\documentclass[sigconf]{acmart}

\usepackage{graphicx}
\usepackage{latexsym}
\usepackage{url}            
\usepackage{booktabs}       
\usepackage{nicefrac}       
\usepackage{microtype}      
\usepackage{latexsym}
\usepackage{xspace}
\usepackage{dsfont}
\usepackage{multirow}
\usepackage{subcaption}
\usepackage{amsmath}
\usepackage{balance}
\usepackage{wrapfig,lipsum}
\usepackage{csquotes}
\usepackage{xcolor}
\usepackage{fancyvrb}
\usepackage[export]{adjustbox}
\usepackage{arydshln}
\usepackage{colortbl}
\usepackage{mdframed}

\usepackage{bm}

\def\eqref#1{equation~\ref{#1}}

\def\1{\bm{1}}

\DeclareMathAlphabet{\mathsfit}{\encodingdefault}{\sfdefault}{m}{sl}
\SetMathAlphabet{\mathsfit}{bold}{\encodingdefault}{\sfdefault}{bx}{n}

\def\sV{{\mathbb{V}}}

\newcommand{\E}{\mathbb{E}}

\DeclareMathOperator*{\argmin}{arg\,min}

\usepackage{array}

\newcolumntype{L}[1]{>{\raggedright\let\newline\\\arraybackslash\hspace{0pt}}m{#1}}
\newcolumntype{C}[1]{>{\centering\let\newline\\\arraybackslash\hspace{0pt}}m{#1}}
\newcolumntype{R}[1]{>{\raggedleft\let\newline\\\arraybackslash\hspace{0pt}}m{#1}}

\makeatletter
\def\adl@drawiv#1#2#3{%
        \hskip.5\tabcolsep
        \xleaders#3{#2.5\@tempdimb #1{1}#2.5\@tempdimb}%
                #2\z@ plus1fil minus1fil\relax
        \hskip.5\tabcolsep}
\newcommand{\cdashlinelr}[1]{%
  \noalign{\vskip\aboverulesep
           \global\let\@dashdrawstore\adl@draw
           \global\let\adl@draw\adl@drawiv}
  \cdashline{#1}
  \noalign{\global\let\adl@draw\@dashdrawstore
           \vskip\belowrulesep}}
\makeatother

\newcommand{\eg}{e.\,g., }
\newcommand{\ie}{i.\,e., }

\newcommand{\modelbm}{\textsc{BM25}\xspace}
\newcommand{\modelprf}{\text{RM3 PRF}\xspace}
\newcommand{\modelmp}{\textsc{MP}\xspace}
\newcommand{\modelknrm}{\textsc{KNRM}\xspace}
\newcommand{\modelconvknrm}{\textsc{C-KNRM}\xspace}
\newcommand{\modeltk}{\textsc{TK}\xspace}
\newcommand{\modelbert}{\textsc{Bert}\xspace}
\newcommand{\modelberttiny}{$\textsc{Bert}_{\text{L2}}$\xspace}
\newcommand{\modelbertmini}{$\textsc{Bert}_{\text{L4}}$\xspace}
\newcommand{\modelbertadv}{\textsc{AdvBert}\xspace}
\newcommand{\modelbertadvtiny}{$\textsc{AdvBert}_{\text{L2}}$\xspace}
\newcommand{\modelbertadvmini}{$\textsc{AdvBert}_{\text{L4}}$\xspace}
\newcommand{\modelrndbm}{$\textsc{Set}_{\text{Top200}}$\xspace}
\newcommand{\modelrndcoll}{$\textsc{Set}_{\text{All}}$\xspace}

\newcommand{\runorig}{\texttt{Orig}\xspace}
\newcommand{\runwqrels}{\texttt{+QRels}\xspace}

\newcommand{\docfactor}{\omega\xspace}
\newcommand{\attmagnitude}{mag\xspace}

\newcommand{\metricndcg}{\text{NDCG}\xspace}
\newcommand{\metricrecall}{\text{Recall}\xspace}
\newcommand{\metricmrr}{\text{MRR}\xspace}
\newcommand{\metricfair}{\text{FaiRR}\xspace}
\newcommand{\metricfairideal}{\text{IFaiRR}\xspace}
\newcommand{\metricfairnorm}{\text{NFaiRR}\xspace}
\newcommand{\metricfairset}{\text{SetFaiRR}\xspace}

\newcommand{\colltrecdlorig}{\text{TRECDL19}\xspace}
\newcommand{\collmsmarco}{$\text{MSMARCO}_{\textsc{Fair}}$\xspace}
\newcommand{\colltrecdl}{$\text{TRECDL19}_{\textsc{Fair}}$\xspace}

\newcommand{\publishurl}{\url{https://github.com/CPJKU/FairnessRetrievalResults}}

\def\vemb{{\mathbf{z}}}

\copyrightyear{2021} 
\acmYear{2021} 
\setcopyright{acmcopyright}\acmConference[SIGIR '21]{Proceedings of the 44th International ACM SIGIR Conference on Research and Development in Information Retrieval}{July 11--15, 2021}{Virtual Event, Canada}
\acmBooktitle{Proceedings of the 44th International ACM SIGIR Conference on Research and Development in Information Retrieval (SIGIR '21), July 11--15, 2021, Virtual Event, Canada}
\acmPrice{15.00}
\acmDOI{10.1145/3404835.3462949}
\acmISBN{978-1-4503-8037-9/21/07}

\begin{document}
\fancyhead{}

\title{Societal Biases in Retrieved Contents: Measurement Framework and Adversarial Mitigation for BERT Rankers}


\author{Navid Rekabsaz}
\email{navid.rekabsaz@jku.at}
\affiliation{%
  \institution{Johannes Kepler University Linz}
  \institution{Linz Institute of Technology, AI Lab}
  \country{Austria}
}

\author{Simone Kopeinik}
\email{skopeinik@know-center.at}
\affiliation{%
  \institution{Know-Center GmbH}
  \country{Austria}
}

\author{Markus Schedl}
\email{markus.schedl@jku.at}
\affiliation{%
  \institution{Johannes Kepler University Linz}
  \institution{Linz Institute of Technology, AI Lab}
  \country{Austria}
}


\begin{abstract}
\vspace{-0.1cm}
Societal biases resonate in the retrieved contents of information retrieval (IR) systems, resulting in reinforcing existing stereotypes. Approaching this issue requires established measures of fairness in respect to the representation of various social groups in retrieval results, as well as methods to mitigate such biases, particularly in the light of the advances in deep ranking models. In this work, we first provide a novel framework to measure the fairness in the retrieved text contents of ranking models. Introducing a ranker-agnostic measurement, the framework also enables the disentanglement of the effect on fairness of collection from that of rankers. To mitigate these biases, we propose \modelbertadv, a ranking model achieved by adapting adversarial bias mitigation for IR, which jointly learns to predict relevance and remove protected attributes. We conduct experiments on two passage retrieval collections (MSMARCO Passage Re-ranking and TREC Deep Learning 2019 Passage Re-ranking), which we extend by fairness annotations of a selected subset of queries regarding gender attributes. Our results on the MSMARCO benchmark show that, (1) all ranking models are less fair in comparison with ranker-agnostic baselines, and (2) the fairness of \modelbert rankers significantly improves when using the proposed \modelbertadv models. Lastly, we investigate the trade-off between fairness and utility, showing that we can maintain the significant improvements in fairness without any significant loss in utility.
\vspace{-0.2cm}
\end{abstract}


\vspace{-0.5cm}
\keywords{fairness, gender bias, neural information retrieval models, BERT, adversarial training, bias mitigation\vspace{-0.1cm}
}


\maketitle

\section{Introduction}
\vspace{-0.1cm}
\label{sec:introduction}
Societal biases and stereotypes are reflected in search engine results, and are most likely reinforced and strengthened by information access systems. The existence of bias in retrieval results, namely the disproportional presence of a specific social group in the contents of retrieved documents, has been shown in several previous studies~\cite{rekabsaz2020neural,kay2015unequal,chen2018investigating,fabris2020gender,otterbacher2017competent}. This topic becomes particularly important considering that users typically tend to accept search engines' results as the ``state of the world''~\cite{pan2007google}. In fact, such biases can lead to representational harms, \ie the representation of some social groups in a less favorable light~\cite{blodgett2020language,rekabsaz2018measuring}, but also to an unfair distribution of opportunities and resources (allocational harms)~\cite{chen2018investigating,biega2018equity}. 

The focus of this work is first on measuring bias/fairness regarding the representation of specific social groups in the retrieval results of IR models, studied on widely-used passage retrieval benchmarks. Our second focus is on the mitigation of these biases by approaching deep retrieval models with an adversarial training method. We conduct our experiments on a human-annotated set of queries, which bias in their retrieval results is considered as \emph{socially problematic}. We refer to such queries as \emph{fairness-sensitive}, where a fair IR system is expected to provide a balanced (or no) representation of the protected attributes (\eg gender, race, ethnicity, and age) in the retrieved contents.

\begin{figure}[t]
\begin{center}
\begin{tabular}{l}
  \small Query: \texttt{how important is a governor?}\\
  \fbox{\parbox{8cm}{\parbox{7.9cm}{
    \scriptsize
    \texttt{The governor is the visible official who commands media attention. The governor, along with the lieutenant governor, is also a major legislative player. [...] The governor has several other important roles. [...] Often overlooked is the role of intergovernmental middleman, a fulcrum of power and a center of political gravity.}}\\
    Ranking position in \modelbertadv: 2
  }}\\

  \fbox{\parbox{8cm}{\colorbox{red!10}{\parbox{7.8cm}{
    \scriptsize
    \texttt{Governor [...] is the chief executive of the state. \textbf{He} is the little president that implements the law in the state and oversee the operations of all local government units within \textbf{his} area. [...] \textbf{He} makes decisions for \textbf{his} state and makes opinions to the people of the state where \textbf{he} is president of the state that \textbf{he} controls[...]}}}\\
    Ranking position in \modelbertadv: 8
  }}\\
  
  \fbox{\parbox{8cm}{\parbox{7.9cm}{
    \scriptsize
    \texttt{The Governor-General is the guardian of the constitution with respect to government ministers, on behalf of the people. This is the most important function of the role. (1)  It ensures the stability of government, irrespective of which political party is in power. (2)  It ensures the legality of government. [...]}}\\
    Ranking position in \modelbertadv: 3
  }}\\

\end{tabular}
\vspace{-0.3cm}
\caption{Top 3 results of a \modelbert ranker on a fairness-sensitive query, retrieved from MSMARCO Passage retrieval. \modelbertadv (our proposed model) ranks the 2nd passage in a lower position.}
\label{fig:problem}
\end{center}
\vspace{-0.7cm}
\end{figure}

Figure~\ref{fig:problem} depicts a representative example in the context of gender bias. The figure shows the top 3 retrieved passages by the \modelbert model~\cite{devlin2019bert}, for a given fairness-sensitive query. Among the retrieved passages, the one in the second ranking position characterizes \emph{governor} (a gender neutral word) as a male, making the overall retrieval results biased. The proposed \modelbertadv model aims to make the relevance prediction of \modelbert invariant to gender features. As shown, \modelbertadv ranks the other two passages still on top of the list, while the biased passage is moved to a lower position in the ranked list. We explain our core contributions in what follows.

In our first contribution, following previous works~\cite{rekabsaz2020neural,fabris2020gender}, we provide a \emph{generic framework to measure fairness of retrieval results} in respect to a protected attribute. The framework consists of three components. The first component measures the neutrality of the content of a document in respect to the protected attribute (\emph{document neutrality}). A document is considered neutral if it contains either no indication, or a balanced representation of the protected attribute. The second component is a novel fairness metric of a ranked list. The metric provides a normalized fairness score, which characterizes how balanced the contents of the ranked list are in regard to the protected attribute. The introduced fairness metric is defined on a given ranked list. As discussed in previous studies~\cite{baeza2018bias,baeza2020bias,ekstrand2019fairness}, various aspects of the IR ecosystem -- \ie collection, model, user feedback, relevance annotation, evaluation, etc. --  can influence  the existence of biases. Among these aspect, the characteristics of a collection can directly affect the results of any chosen IR model and is therefore central in fairness measurement.\footnote{Consider the extreme case of a collection in which all the documents that contain the term \emph{nurse}, refer to it as a female. In this case, regardless of the choice of the IR model, the retrieved \emph{nurse}-related documents provide biased results towards female.} The third component of our framework therefore aims to disentangle the effect of the collection on the fairness of retrieval results, from the one of an IR model. To this aim, we introduce a ranker-agnostic fairness metric, defined as the expectation over the fairness of all possible ranking permutations of a set of documents. The resulting metric is cheap to compute, and solely reflects the characteristics of the given document set by factoring out the effect of retrieval models on the fairness metric.

As a second contribution, we introduce the \textit{adversarial bias mitigation method} to deep neural ranking models, drawing from the literature of learning invariant representations~\cite{madras2018learning,zhang2018mitigating,ganin2015unsupervised,elazar2018adversarial,xie2017controllable}. In our proposed method, the \modelbert ranker is extended with an adversarial training mechanism, which aims to make the relevance scoring of the model invariant to the protected attribute. The adversarial method is jointly optimized with the model's main objective, aiming to maintain the effectiveness of relevance prediction (characterized by utility evaluation scores), while reducing bias. 

To enable studying the fairness of retrieval results on public collections, in the third contribution, we provide \emph{two novel datasets of fairness-sensitive queries in respect to gender}. The datasets of queries are chosen from the sets of queries of the MSMARCO Passage Re-ranking collection~\cite{nguyen2016ms} and TREC Deep Learning Track 2019 Passage Retrieval task~\cite{craswell2020overview}, and referred to as \collmsmarco and \colltrecdl, respectively. The datasets are created by human annotators, where each query is judged as being essential for the study of gender fairness in retrieval results. These datasets facilitate the research on fairness together with utility of IR models, and are especially suited for the studies of deep ranking models. 


Using these gender fairness-sensitive queries, in the last contribution, we conduct a \emph{large set of experiments} on various classical and neural/deep IR models on the \collmsmarco and \colltrecdl collection. We in particular study the effect of the adversarial training method in \modelbertadv. The evaluation results on our introduced fairness metric as well as several utility metrics show that, on the \collmsmarco collection all IR models have a lower fairness score than the ones of ranker-agnostic document sets. This highlights the potentials for improving the fairness of the ranking models. In particular, we observe a significant improvement in the fairness of \modelbertadv in comparison with \modelbert, indicating the effectiveness of the adversarial training method. The results on the \colltrecdl collection shows that the fairness scores of the selected queries are already close to maximum, while \modelbertadv similarly improves the fairness of \modelbert. Finally, by introducing a fairness--utility trade-off approach, we show that by correctly selecting a version of \modelbertadv, we can achieve significant improvement in the fairness metric with no significant loss on performance.


The paper is structured as follows: Related work is discussed in Section~\ref{sec:related}. Our fairness measurement framework is then detailed in Section~\ref{sec:measurement}. The procedure of creating the datasets is explained in Section~\ref{sec:collection}, and the adversarial bias mitigation method in Section~\ref{sec:adversarial}. Section~\ref{sec:setup} describes the experiments, whose results are presented and discussed in Section~\ref{sec:results}. The source code together with all resources used in this study are available at \textbf{\publishurl}.

\vspace{-0.3cm}
\section{Related Work}
\label{sec:related}
Various forms of societal biases are involved in the ecosystem of information systems~\cite{baeza2018bias,baeza2020bias}. \citet{ekstrand2019fairness} point out various sources of such biases, \ie data collection, model, evaluation, and human interaction. Among these, the focus of the current study is on model and collection. 

Bias and fairness is explored in various IR scenarios. \citet{geyik2019fairness} explore fairness and bias in a large-scale talent search platform, while \citet{chen2018investigating} look at individual and group unfairness in the results of commercial resume search engines. Fairness in the matching process of two-sided markets is approached by \citet{suhr2019two}, and later by \citet{morik2020controlling} who propose a dynamically adapting controller to integrate both fairness and utility. \citet{otterbacher2018investigating} investigate the perception and prejudices of users when interacting with an image search engine, and finally, \citet{gerritse2020bias} frame the issue of biases in the context of personalization in conversational search. 

Regarding bias and fairness measurement in ranked lists, \citet{singh2018fairness} formulate the group fairness in rankings in terms of exposure allocation, and introduce various optimization constraints to satisfy fairness in sense of demographic parity, disparate treatment, and disparate impact. Our proposed framework contributes to this direction by proposing a metric of group fairness, which extends the concept of demographic parity by including the cases that do not contain any protected attribute. Complementing the studies on group fairness, \citet{biega2018equity} define a metric for individual fairness based on the notion of amortized fairness, where the accumulated attention to individual items across a set of rankings should be proportional to the accumulated relevance. 

A few works have studied the existence of bias in the contents of retrieval results. \citet{kay2015unequal} evaluate the presence of gender bias in image search results for a variety of occupations, showing the reinforcement of stereotypes towards and the systematic under-representation of women in search results. More recently, \citet{fabris2020gender} quantify the reinforcement of gender stereotypes in retrieval results, and conduct experiments on synthetic and a standard benchmark using classical and word embedding-based IR models. \citet{gao2020toward} focus on fairness regarding the diversity of the topics which appear in retrieval results, measured on a commercial search engine as well as standard TREC benchmarks. Following this direction, \citet{rekabsaz2020neural} explore the degree of gender bias in the retrieved passages by several neural IR models from a set of non-gendered queries, observing the effects of learned and transferred embeddings on this form of bias. The present work directly contributes to this line of research by providing a framework for measuring bias in retrieved contents as well as two sets of fairness-sensitive queries, which facilitate further reproducible research.  

The approaches to bias mitigation in deep learning are categorized into pre-processing, in-processing, and post-processing~\cite{mehrabi2019survey}. The current study focuses on in-processing approaches, namely by adapting the model to satisfy fairness as well as utility objectives. Other in-processing approaches include \citet{singh2019policy}, who propose a generic fairness-aware learning to rank (LTR) framework by introducing a policy-gradient approach to impose fairness constraints in a listwise LTR setting. More recently, \citet{zehlike2020reducing} approach the integration of fairness in listwise LTR through a regularization term added to the model's utility objective. Our work contributes to this research direction by applying adversarial training to a pairwise LTR setting.

Regarding post-processing approaches, several studies propose methods to minimize the representational differences between the groups in a ranking list~\cite{zehlike2017fa,celis2018ranking,yang2017measuring}. Finally, the pre-processing approaches focus on balancing or manipulating the collection or training data. As an example, \citet{de2019bias} scrape the gender-related words in a set of biographies, and observe considerable improvement in the fairness of a classifier that predicts the corresponding occupations. In a pilot study, we also experiment with scraping gender-related words from query and document, but do not observe any improvement in the fairness of the resulting ranked list. The focus of the current work therefore remains on in-processing bias mitigation approaches, particularly through adversarial training.

Learning \textit{fair representations} is first introduced by \citet{zemel2013learning}. The goal of their proposed method is to achieve embeddings which simultaneously provide a good encoding of data, while removing any information about protected attributes. Related to learning invariant representations, \citet{ganin2015unsupervised} propose adversarial training for domain adaptation, where the learned feature embeddings should be discriminative for the main task while indiscriminate towards the shifts between domains. Following this study, \citet{xie2017controllable} and later \citet{madras2018learning} introduce adversarial learning to the context of fair representation learning. Recently, \citet{elazar2018adversarial} and \citet{barrett2019adversarial} investigate the use of adversarial training for removing demographic information from the intermediary embeddings of a neural text classifier. Our work is closely related to these studies by introducing adversarial training to pairwise LTR and \modelbert ranking models.






\vspace{-0.2cm}
\section{Fairness in Retrieval Results}
\label{sec:measurement}
We now explain our novel framework to measuring the degree of fairness in retrieval results, namely to what extent the contents of the retrieved documents picture a balanced representation in respect to a protected attribute (such as gender, race, ethnicity, age, etc.). We first explain the approach to calculating the neutrality of a document. Using this measure of document neutrality, we next introduce a metric to quantify the fairness in a ranked list of documents. Considering these definitions, we finally explain the method to calculate the fairness of a subset of documents in collection, which is independent of the chosen ranking model. The introduced measurements are generic and flexible, and can be applied to any definition of protected attributes. The framework assumes the existence of a set of fairness-sensitive queries $Q$, whose results are expected to be balanced across the members of a protected attribute, denoted by the set $A$. We will provide a set of such queries for genders in Section~\ref{sec:collection}. 

\vspace{-0.2cm}
\subsection{Document Neutrality}
A document is considered as neutral if it either does not contain any indication of none of the members of the protected attribute, or if the document contains a balanced representation of those members. To formally define the latter in a generic way, we first introduce the random variable $J$, which indicates the expected proportion of each protected member in a fully balanced/fair representation. Concretely, for each member $a \in A$, the corresponding value of $J$, $J_a$ should be defined according to what we expect as a balanced document, such that $\sum_{a \in A}{J_{a}=1}$. For instance, $J$ in a binary setting for genders can be defined as $J_{female}=1/2$, and $J_{male}=1/2$. The definition of $J$ is generic and can cover subtle definitions beyond binary assumptions.

Following the common practices in the studies of bias in text~\cite{caliskan2017semantics,rekabsaz2020neural,rekabsaz2018measuring,bolukbasi2016man}, we define each member of the protected attribute with the set of words $\sV_a$. These words are strong indicatives of the member $a$, which we refer to as \emph{representative words}.\footnote{For instance, words such as \emph{she}, \emph{woman}, \emph{her} are used to define female, and \emph{he}, \emph{man}, \emph{him} to define male.} The use of these small sets of representative words indeed does not cover all the incidents related to the protected attribute of interest. Though, this can be seen as a precision-oriented approach, which aims to avoid the introduction of noise to the approximated quantities.

We now define the magnitude of existence of each protected attribute's member in a document, $\attmagnitude^{a}\!(d)$, as the sum of the number of occurrences of each word in $\sV_{a}$ in the document, formulated as follows: 
\begin{equation}
\attmagnitude^{a}(d)=\sum_{w\in\sV_a}{\#\langle w, d\rangle}
\label{eq:measurement:f}
\end{equation}
where $\#\langle w, d\rangle$ indicates the number of times the word $w$ appears in~$d$. A similar quantity is used by \citet{rekabsaz2020neural}. Based on this definition, the neutrality of document $d$, $\docfactor(d)$, is defined as follows:
\begin{equation}
\docfactor(d)=\begin{cases}
    1,& \text{if } \sum_{a \in A}{\attmagnitude^{a}\!(d)} \leq \tau\\
    1 - \sum_{a \in A}{\left\lvert \frac{\attmagnitude^{a}\!(d)}{\sum_{x \in A}{ \attmagnitude^{x}\!(d) }} - J_{a}\right\rvert},& \text{otherwise}
\end{cases}
\label{eq:measurement:docfactor}
\end{equation}
where $\tau$ denotes the threshold parameter on the sum of the magnitudes of all members, below which the document is considered as neutral. The threshold $\tau$ is introduced to reduce the effect of noise by drawing a (hard) line between the documents, considered as neutral versus the non-neutral ones. The possible values of $\docfactor(d)$ are always between 0 and 1, where 1 shows the full neutrality of the document, and 0 indicates the dominant existence of one of the members of the protected attribute. For instance, in the case of $J_{female}=1/2$ and $J_{male}=1/2$, if a document has no representative words, or an equal number of occurrences of female- and male-representative words, $\docfactor(d)$ is equal to 1. If only one gender is represented in the document, $\docfactor(d)$ is equal to 0. If both genders appear but in an unbalanced way, $\docfactor(d)$ is between 0 and 1.\footnote{As a more nuanced example, if $\attmagnitude(d)$ values for female and male are 6 and 4 respectively, $\docfactor(d)=0.8$.}

\vspace{-0.2cm}
\subsection{Fairness of Ranked Lists}
Using document neutrality, we now introduce a metric to measure the fairness of the retrieved documents, given a query. This measure takes into account the neutrality of every document in the ranked list, but also the importance of the position of each document in the list (position bias). Similar to previous studies~\cite{singh2018fairness,fabris2020gender,kulshrestha2017quantifying}, we define the importance of each position with the function $p(i)$, which monotonically decreases as the position $i$ increases. 

Considering these, the \emph{Fairness of Retrieval Results (\metricfair)} of query $q$ for a set of ranked lists $L$ is defined as follows:
\begin{equation}
\metricfair_{q}\!\left(L \right)=\sum_{i=1}^{t}{\docfactor\!\left(L_{i}^{q}\right)p(i)}
\label{eq:measurement:metricfair}
\end{equation}
where $L^{q}$ is the ranked list of $q$, $t$ is the cut-off threshold on the ranked list, and $L_{i}^{q}$ denotes the document at position $i$ of the ranked list $L^{q}$. The position bias is set to $p(i)=\frac{1}{\log_{2}(1+i)}$ just like in standard definition of Discounted Cumulative Gain (DCG)~\cite{jarvelin2002cumulated} as well as in previous studies~\cite{singh2018fairness,zehlike2020reducing}.

One important consideration for this fairness metric is that -- based on the distribution of document neutrality in the collection -- different queries may end up in different value ranges, and hence might not be directly comparable. To avoid this issue, we introduce a normalization step to $\metricfair_{q}$, by following a similar principle to the one in Normalized DCG (\metricndcg).

To this end, we first consider a set of potentially relevant documents to the query $q$, characterized by the set of documents at the top of the ranking list $L^{q}$ (\eg at top 200 or 1000). We refer to this as \emph{background document set}, and denote it with $\widehat{S}^{q}\subset C$, where $C$ is the set of all documents in collection. In fact, $L^{q}$ is one way of ranking the documents in $\widehat{S}^{q}$.\footnote{Such sets of documents are commonly defined and exploited in various IR tasks. For instance, in the reranking approach using the top results of a first-stage ranker, or in the selection of candidate documents for relevance judgement~\cite{zobel1998reliable,sparck1975report}.}\footnote{The size of $\widehat{S}^{q}$ is typically much smaller than $|C|$. However, with no loss of generality, the definition of background document set can be reduced to the set of all document in collection ($\widehat{S}^{q}=C$).}

Using $\widehat{S}^{q}$, we introduce \emph{Ideal \metricfair ($\metricfairideal$)}, defined for a query $q$ as the best possible fairness result that can be achieved from reordering the documents in $\widehat{S}^{q}$. More specifically, $\metricfairideal_{q}(\widehat{S})$ is computed by sorting the neutrality scores of the documents in $\widehat{S}^{q}$ in descending order, and calculating the \metricfair of the resulted ranked list (according to Eq.~\ref{eq:measurement:metricfair}).

We use \metricfairideal to normalize \metricfair, resulting in \emph{Normalized Fairness of Retrieval Results (\metricfairnorm)} for a given set of ranked lists $L$:
\begin{equation}
\metricfairnorm_{q}\!\left(L,\widehat{S}\right)=\frac{\metricfair_{q}\!\left(L\right)}{\metricfairideal_{q}\!\left(\widehat{S}\right)} 
\label{eq:measurement:metricfairnorm}
\end{equation}

Finally, given the set of queries $Q$, \metricfairnorm of an IR model is defined as the average of per-query scores: 
\begin{equation}
\metricfairnorm(L,\widehat{S})=\sum_{q \in Q}{\metricfairnorm_{q}(L,\widehat{S})}
\label{eq:measurement:metricfairnorm_model}
\end{equation}

The possible values for \metricfairnorm range between 0 and 1, where 0 indicates the maximum amount of bias/unfairness, and 1 the maximum possible fairness in the retrieval results.

\vspace{-0.2cm}
\subsection{Ranker-Agnostic Fairness of Document Sets}
The proposed $\metricfairnorm_{q}$ is so far defined as a fairness metric for the given ranked list $L^{q}$. As mentioned in Section~\ref{sec:introduction}, we are also interested in measuring the fairness on a set of documents, while excluding the effect of any particular ranker (a ranker-agnostic metric). More formally, given a set of documents $S$, we aim to calculate the fairness metric for the set, independent of any possible ranking permutation that can be created from the documents in $S$. We refer to this set for a specific query as $S^{q}$. In principle, $S^{q}$ can be defined as any set of documents, such as the background set ($S^{q}=\widehat{S}^{q}$), or whole the collection ($S^{q}=C$).

To provide such ranker-agnostic fairness metric, we first define $\Psi\!(S^{q})$ as the set of all possible ranking permutations that can be created for the set of documents $S^{q}$.\footnote{The size of the corresponding $\Psi$ set is $|\Psi\!(S^{q})|=|S^{q}|!$} Using $\Psi$, we define \metricfairset as the expectation of the \metricfair quantity over all ranking permutations, defined as follows:
\begin{equation}
\metricfairset_{q}\!\left(S\right)=\mathop{\E}_{L\sim \Psi\!(S^{q})}{\left[ \metricfair_{q}(L) \right]}
\label{eq:measurement:metricfairset}
\end{equation}

The applied expectation over $\Psi$ in fact factors out the effect of any specific ranker, and therefore \metricfairset solely quantifies the degree of fairness of the $S^{q}$ set. By putting the definition of \metricfair from Eq.~\ref{eq:measurement:metricfair} into Eq.~\ref{eq:measurement:metricfairset}, and by considering that the position bias $p(i)$ is invariant to the choice of the ranker, we achieve:
\begin{equation}
\begin{gathered}
\metricfairset_{q}\!\left(S\right)=\mathop{\E}_{L\sim \Psi\!(S^{q})}{\left[ \sum_{i=1}^{t}{\docfactor\!\left(L_{i}^{q}\right)p(i)} \right]}=\\
\sum_{i=1}^{t}{\mathop{\E}_{L\sim \Psi\!(S^{q})}{\left[\docfactor\!\left(L_{i}^{q}\right)\right] p(i)}}
\end{gathered}
\end{equation}

In the equation above, the term $\mathop{\E}_{L\sim \Psi\!(S^{q})}{[\docfactor\!(L_{i}^{q})]}$ is in fact equivalent to the expectation of the neutrality score of any document in $S^{q}$, that appear in the position $i$. The value of the mentioned quantity is the same for any position and is equal to the expectation of the neutrality scores of the documents in $S^{q}$, formulated below:
\begin{equation}
\mathop{\E}_{L\sim \Psi\!(S^{q})}\left[\docfactor\!\left(L_{i}^{q}\right)\right]=
\mathop{\E}_{d\sim S^{q}}\left[ \docfactor\!\left(d\right) \right]=\frac{\sum_{d\in S^{q}}{\docfactor\!\left(d\right)}}{\lvert S^{q}\rvert}
\label{eq:measurement:metricfairset_expectations}
\end{equation}

In fact, to calculate $\mathop{\E}_{L\sim \Psi(S^{q})}{[\docfactor(L_{i}^{q})]}$, we do not need to create all possible permutations, but can simply compute the mean of the neutrality scores of the documents in $L^{q}$. Putting the results of Eq.~\ref{eq:measurement:metricfairset_expectations} into Eq.~\ref{eq:measurement:metricfairset}, we achieve the final definition of \metricfairset, shown below:
\begin{equation*}
\begin{gathered}
\metricfairset_{q}\!\left(S\right)=\sum_{i=1}^{t}{\frac{ \sum_{d\in S^{q}}{\docfactor\!\left(d\right)} }{\lvert S^{q}\rvert}p(i)} =
\frac{t \times \sum_{d\in S^{q}}{\docfactor\!\left(d\right)}}{\lvert S^{q}\rvert} \sum_{i=1}^{t}{p(i)} 
\end{gathered}
\end{equation*}
where as before $t$ is the cutoff threshold. Similar to \metricfair, we normalize this ranker-agnostic fairness metric using \metricfairideal, as defined below:
\begin{equation}
\metricfairnorm_q\left(S,\widehat{S}\right)=\frac{\metricfairset_{q}\!\left(S\right)}{\metricfairideal_{q}\!\left(\widehat{S}\right)} 
\label{eq:measurement:metricfairsetnorm}
\end{equation}

The ranker-agnostic \metricfairnorm for a document set is similarly calculated as the average of $\metricfairnorm_q$ values over the query set $Q$ (Eq.~\ref{eq:measurement:metricfairnorm_model}). We should emphasize that the ranker-agnostic \metricfairnorm values are indeed directly comparable with the \metricfairnorm values calculated on specific ranked lists, as far the metrics are normalized with the same background document sets $\widehat{S}$. In other words, $\metricfairnorm_q(S,\widehat{S})$ can be seen as the results of a random ranker, where the quantified measure only reflects the characteristic of $S$. Another consideration regarding $\metricfairnorm_q(S,\widehat{S})$ is that, in contrast to $\metricfairnorm_q(R,\widehat{S})$, it can have values higher than 1, since we did not limit the definition of $S$ to be the same as $\widehat{S}$. However, in our experiments, we consistently only observe values smaller than 1.

\vspace{-0.2cm}
\section{Fairness-Sensitive Queries Dataset}
\label{sec:collection}
We create two datasets of fairness-sensitive queries in respect to gender equality, namely \collmsmarco and \colltrecdl. These datasets form a subset of the queries of the TREC Deep Learning Track 2019 Passage Retrieval (\colltrecdlorig)~\cite{craswell2020overview} and the development set of the MSMARCO Passage Reranking~\cite{nguyen2016ms}. \citet{rekabsaz2020neural} provide 1,765 non-gendered queries, as a subset of 55,578 queries of MS~MARCO, which are annotated by three Amazon Mechanical Turk workers. The annotators were asked to mark each query that contains at least one word or phrase that refers to gender-related concepts. Following this approach, we annotate the queries of \colltrecdlorig similarly with three Amazon Mechanical Turk workers (English native-speakers). Then, we select from both datasets the queries annotated as non-gendered by the majority of workers. The resulting two sets of queries form the starting point for a meta-annotation that further reduces the queries to a more focused subset.

The aim of the meta-annotation is firstly to verify the annotations of the workers, but also to mark the queries, for which the existence of gender bias in their retrieval results is considered as \emph{socially problematic}. In particular, the meta-annotators first repeat the effort of the crowd workers and ask ``is the query non-gendered?''. Then, the subject matter becomes central as they inquire ``is the existence of bias in retrieval result socially problematic?''. 

\begin{table}[t]
\begin{center}
\caption{Samples of fairness-sensitive queries.}
\vspace{-0.3cm}
\small
\begin{tabular}{p{0.02cm} p{7.7cm}}%
\toprule
\multicolumn{2}{l}{Query: \texttt{how important is a governor?}}\\
& Domains: Career, Politics \\
& Reasoning: Bias contributes to existing stereotypes of career choices.\\
\multicolumn{2}{l}{Query: \texttt{when do babies start eating whole foods}}\\
& Domains: Social inequality, Career \\
& Reasoning: Bias contributes to the gender norm "women as a caretaker", and consequently might impact career choices.\\
\multicolumn{2}{l}{Query: \texttt{how do i figure my normal bmi}}\\
& Domains: Health\\
& Reasoning: Bias suggests that it is more important for women to maintain their weight. One implication might be the affirmation of conventional norms of strong masculinity, which results in men being more likely to live unhealthy and consume harmful substances.\\
\bottomrule
\end{tabular}
\label{tbl:collection:queries} 
\end{center}
\vspace{-0.7cm}
\end{table}

Let us discuss more concretely what \emph{socially problematic} refers to in the context of this meta-annotation. The aim of the annotation is to identify fairness-sensitive queries that in case of biased search results, potentially reinforce existing gender norms and thus promote gender inequality. While gender roles have become more flexible within the last century, specific expectations on how men and women have to act still exist in today's society. Socialization, driven by the impact of parents, peers, and media, is a crucial factor in the individual's learning of gender roles~\cite{wienclaw2011gender}. Likewise, online information and how it is biased towards gender contributes to the individual and social understanding of such gender roles and, as a consequence, manifestations of gender inequality. Thus, within this context, \emph{socially problematic} refers to all queries that relate to a selected list of domains that are recognized to face challenges in achieving gender equality~\cite{lorber2005breaking,UNSDG}. In the conducted meta-annotation, this list of domains includes Education (e.g., degree of education, career choice), Career (e.g., gender pay gap, labor force participation), Health (e.g., toxic masculinity), Violence and Exploitation, Social Inequality (e.g., domestic duties, access to justice, access to finance and property), and Politics (e.g., power representation). Table \ref{tbl:collection:queries} shows three examples of fairness-sensitive queries, their assignment to a domain, and related reasoning.




The meta-annotation is performed by two post-doctoral researchers individually. Both are experts in research on biases and computer science. The final two datasets of fairness-sensitive queries, \collmsmarco and \colltrecdl, contain only those items that both meta-annotators agree on to be \emph{non-gendered} and \emph{socially problematic}. These two datasets include a total of \textbf{215} and \textbf{30 queries} for \collmsmarco and \colltrecdl, respectively. Please note that we consider this an initial set and understand the extension of impact domains and their assignment to queries as a collaborative effort within the research community. To facilitate this, together with the query sets, we also make public the corresponding domains and reasoning. 

We should finally highlight a practical difference between these two sets. The queries in \collmsmarco are shortlisted from a large set of queries, and the resulting dataset provides a challenging benchmark for studying fairness in retrieved contents. In contrast, the dataset of \colltrecdl is selected from the much smaller set of queries in \colltrecdlorig, and studying it mainly examines the characteristics of a common IR evaluation benchmark through the lens of fairness in retrieved contents.

\vspace{-0.2cm}
\section{Adversarial Bias Mitigation}
\label{sec:adversarial}
This section describes the proposed \modelbertadv model, which extends the \modelbert ranker model with adversarial training. We introduce this adversarial mechanism into the architecture of \modelbert due to \modelbert's impressive performance for retrieval tasks~\cite{nogueira2019passage}, and also as this model has been the basis for several recent ranking models~\cite{macavaney2020efficient,xiong2021approximate,khattab2020colbert}. The proposed methodology can readily be adopted to neural models other than the basic form of \modelbert, such as other \modelbert-based variations~\cite{macavaney2020efficient} or dense retrieval approaches~\cite{xiong2021approximate,khattab2020colbert}. 

We follow the basic setup of pairwise LTR in which each data point in the given training data consists of a pair of a positive and a negative data item, $\langle q,X^{+},X^{-}\rangle$, where $X^{+}$ and $X^{-}$ refer to the set containing the information about a relevant and a non-relevant document to $q$, respectively. Each of $X^{+}$ and $X^{-}$ has two elements: $\langle d^{+},l^{+}\rangle$ for the former and $\langle d^{-},l^{-}\rangle$ for the latter, where $d$ refers to the corresponding document, and $l$ denotes the corresponding label regarding the protected attribute. We define $l$ as a binary variable, indicating whether the combination of $q$ with the corresponding $d$ contains the protected attribute. Concretely, the protected label $l$ is equal to 1 if the concatenation of $q$ and $d$, $[d,q]$, is a neutral piece of text, namely when $\docfactor\left([d,q]\right)<1$ (see Eq.~\ref{eq:measurement:docfactor}), and 0 otherwise. We next explain the architecture of the \modelbertadv model, followed by describing the adversarial training method.

\vspace{-0.2cm}
\subsection{\modelbertadv Model}
A \modelbert ranker receives $q$ and $d$ as input, and encodes them to an interaction embedding vector $\vemb$ (resulted as the direct output of the $\texttt{[CLS]}$ special token). We refer to $\vemb$ as query-document interaction embedding, and denote this process with the encoding function $f$ such that $\vemb=f(q,d)$. \modelbert uses $\vemb$ to predict the relevance of $q$ to $d$ through the neural function $g(\vemb)$, where the output relevance score is predicted by $g(f(q,d))$. The function $g$ is defined as a linear projection of $\vemb$ to a relevance score. We refer to this combination of $f$ and $q$ as \emph{utility network}.

\begin{figure}[t]
\centering
\includegraphics[width=0.25\textwidth]{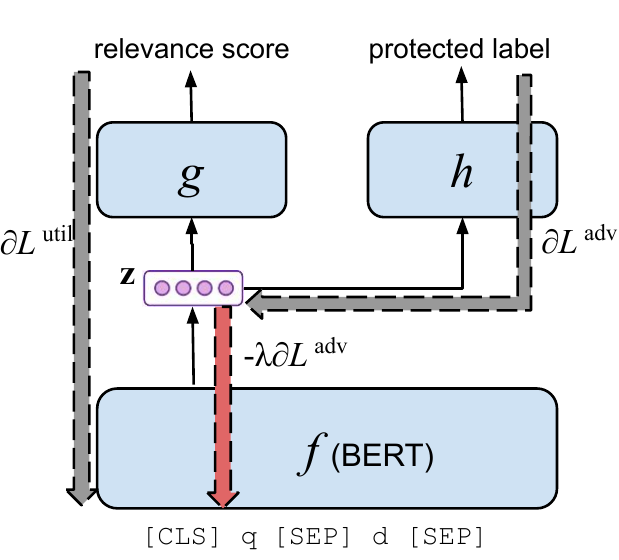}
\centering
\vspace{-0.1cm}
\caption{The schematic view of \modelbertadv. The red arrow shows the reverse gradient from the adversarial network.}
\label{fig:adversarial}
\vspace{-0.5cm}
\end{figure}

Following previous work~\cite{ganin2015unsupervised,xie2017controllable,elazar2018adversarial}, the proposed \modelbertadv defines an additional classifier head on top of $\vemb$, which aims to predict the protected label from the encoded information in the query-document interaction embedding. The adversarial classifier is defined as the neural function $h(\vemb)$, which is a two-layer feedforward network with a tanh non-linearity followed by a softmax layer. The output of the \emph{adversarial network} is therefore achieved through the function $h(f(q,d))$. A schematic view of the model is shown in Figure~\ref{fig:adversarial}. 

\vspace{-0.2cm}
\subsection{Adversarial Training}
The objective of adversarial training is to make the interaction embedding $\vemb$ invariant to the protected attribute, namely to the prediction of the protected label. In other words, we aim to learn the encoder $f$ in such a way that its output embedding $\vemb$ is minimally informative for predicting $l$, while simultaneously it is maximally informative for predicting relevance scores. Following the adversarial training setup~\cite{goodfellow2014generative,ganin2015unsupervised,xie2017controllable,elazar2018adversarial}, the adversarial classifier $h(\vemb)$ is therefore trained to predict $l$, while the encoder $f$ is trained to make $h(\vemb)$ fail. This mechanism is defined in a min-max game, where the network tries to jointly optimize these two objectives. For a data point $\langle q,X^{+},X^{-}\rangle$, the overall objective $\mathcal{L}$ as defined as follows:
\begin{equation}
\vspace{-0.1cm}
\begin{gathered}
\arg\min_{f,g}\max_{h} \mathcal{L}=\mathcal{L}^{\text{util}}\left(q,X^{+},X^{-}\right) - \mathcal{L}^{\text{adv}}\left(q,X^{+}\right) - \mathcal{L}^{\text{adv}}\left(q,X^{-}\right)\\
\begin{aligned}
\mathcal{L}^{\text{util}}\left(q,X^{+},X^{-}\right)=&\mathcal{L^{\text{MM}}}{\left(g(f(q,d^{+})),g(f(q,d^{-}))\right)}\\
\mathcal{L}^{\text{adv}}\left(q,X\right)=&\mathcal{L^{\text{CE}}}{\left(h(f(q,d)),l\right)} 
\end{aligned}
\end{gathered}
\label{eq:adversarial:minmax}
\vspace{-0.1cm}
\end{equation}
where $\mathcal{L}^{\text{util}}$ and $\mathcal{L}^{\text{adv}}$ denote the loss function of the utility and adversarial network, respectively, and $X$ is a generic identifier for either $X^{+}$ or $X^{-}$. The utility network is typically optimized using the max-margin (hinge) loss function on the predictions of the positive and negative document, denoted by $\mathcal{L}^{\text{MM}}$. A variation of this optimization in IR is the sum of the cross entropy loss values of the positive and negative document, where the loss is defined on the two-dimensional (relevant/non-relevant) probability distribution output vector. Regardless of the optimization variations, the adversarial network applies the cross entropy loss $\mathcal{L}^{\text{CE}}$ calculated separately for $X^{+}$ and~$X^{-}$, namely for $\langle d^{+},l^{+}\rangle$ and $\langle d^{-},l^{-}\rangle$.

To optimize this network as suggested by \citet{ganin2015unsupervised}, \modelbertadv uses the \emph{gradient-reversal layer} (GRL), inserted as the layer $rev_{\lambda}$ between the encoded embedding $\vemb$ and the adversarial classifier~$g$. The layer $rev_{\lambda}$ acts as the identity function during the forward pass, while during backpropagation it multiplies the passed gradients by a factor of $-\lambda$. This results in no change in the gradient of $g$, but receiving the gradient of the adversary in the opposite direction to the encoder. The scale of this reversed gradient is set by the hyper-parameter $\lambda$. This reversion in gradient through GRL is depicted in Figure~\ref{fig:adversarial} with the red arrow. Adding GRL to the network in fact simplifies the learning process to a standard gradient-based optimization, in which $\mathcal{L}$ and $\mathcal{L}^{\text{adv}}$ are reformulated as follows:
\begin{equation}
\begin{gathered}
\argmin_{f,g,h} \mathcal{L}=\mathcal{L}^{\text{util}}\left(q,X^{+},X^{-}\right) + \mathcal{L}^{\text{adv}}\left(q,X^{+}\right) + \mathcal{L}^{\text{adv}}\left(q,X^{-}\right)\\
\mathcal{L}^{\text{adv}}\left(q,X\right)=\mathcal{L^{\text{CE}}}{\left(h(rev_{\lambda}(f(q,d))),l\right)} 
\end{gathered}
\label{eq:adversarial:final}
\end{equation}

We should note that the adversarial learning does not directly optimize for the fairness metric, but aims to reduce the information regarding the protected attribute in the query-document interaction embedding. We hypothesize that, by providing relevance scores that are maximally independent of the protected attribute through this adversarial training, we can achieve ranking results with less bias in contents. We will test this hypothesis through experimental evaluation in Section~\ref{sec:results}.

\vspace{-0.2cm}
\section{Experiment Setup}
\label{sec:setup}
\vspace{-0.1cm}
\paragraph{Resources}
The fairness and performance of the models are evaluated on the fairness-sensitive queries provided by \collmsmarco and \colltrecdl. Both query datasets share the same document collection, i.e., the MSMARCO Passage Retrieval collection, which contains 8,841,822 passages.

The protected attribute in our experiments is gender, defined in a binary form such that $A=\left\{\text{female},\text{male}\right\}$. The decision of simplifying gender as a binary construct is taken due to practical constraint. We however acknowledge that this choice neglects the broad meaning of gender, and is not representative of all individuals. To define each gender, we use the sets of gender-representative words introduced in previous work~\cite{rekabsaz2020neural,caliskan2017semantics}. In addition, considering the importance of names in gender bias measurement, shown in previous studies~\cite{romanov2019s,maudslay2019s}, we enrich the list of gender-representative words with a focused set of names. This set of names is created based on the dataset of names and their corresponding statistics in the United States population, provided by Social Security Administration (SSA) dataset.\footnote{\url{https://www.ssa.gov/oact/babynames/background.html}} From this dataset, we select an equal number of names for female and male, such that each selected name is assigned to a female or male in at least 75\% of births cases. This additional set enriches the original gender-representative words, while still maintains high precision in defining the protected attribute. The final set defines each gender with 158 words.

\vspace{-0.2cm}
\paragraph{IR Models} 
We conduct our experiments on the following neural IR models: Match Pyramid (\modelmp)~\cite{pang2016text}, Kernel-based Neural Ranking Model (\modelknrm)~\cite{xiong2017end}, Convolutional KNRM (\modelconvknrm)~\cite{dai2018convolutional}, Transformer-Kernel (\modeltk)~\cite{Hofstaetter2020_sigir}, and the fine-tuned \modelbert model~\cite{devlin2019bert}. In addition, we investigate classical IR models, namely \modelbm~\cite{robertson2009probabilistic} and \modelprf~\cite{abdul2004umass}. The \modelbm and \modelprf models are computed using the Anserini~\cite{yang2017anserini} toolkit. The neural models except the \modelbert rankers are trained with the same setting as suggested by \citet{rekabsaz2020neural}. For \modelbert rankers, we investigate two recently-released versions of pre-trained language models known as BERT-Tiny, and BERT-Mini~\cite{turc2019wellread}. The BERT-Tiny, and BERT-Mini models consist of two and four layers of Transformers, respectively, and therefore we refer to the ranker models based on these as \modelberttiny, and \modelbertmini. These \modelbert rankers are fine-tuned according to the training setting suggested by \citet{nogueira2019passage}. 



We use the provided training data of the MSMARCO collection to train the neural IR models. We apply early stopping by following the method suggested by \citet{hofstatter2019effect}, while avoiding any overlap in the provided validation set with the fairness-sensitive queries. Following \citet{rekabsaz2020neural}, the neural models rerank the top 200 retrieval passages of the \modelbm model. The complete parameter settings of all models are provided in the published repository together with the resources and source code.\footnote{\publishurl}


\vspace{-0.2cm}
\paragraph{Document Sets for Studying Ranker-Agnostic Fairness} We investigate the ranker-agnostic fairness regarding two sets of documents. The first set considers all the documents in the collection for any query ($S^{q}=C$). This set is referred to as \modelrndcoll, and aims to reveal the overall characteristics of the collection regarding the representation of genders in the collection. In the second set, the assigned documents for each query are are taken from the top 200 passages retrieved by the \modelbm model. We refer to this set as \modelrndbm. The \modelrndbm set is in fact identical to the document sets used by the neural models for reranking. 

\vspace{-0.2cm}
\paragraph{Oracle Ranking List Setting} In addition to the discussed settings, we are interested in examining the fairness of the models in the hypothetical scenario, where a model provides its best possible ranking according to retrieval utility criteria by using the provided relevance judgment. To this end, we create a new variation of any given ranked list, in which the relevant passages (in QRels) are moved to the top of the list. We refer to the actual ranked lists as \runorig, and to the variations with this oracle knowledge as \runwqrels.

\vspace{-0.2cm}
\paragraph{Adversarial Training Procedure}
To train the adversarial network introduced in Section~\ref{sec:adversarial}, we assign a gendered/non-gendered label to each data point of the training data. Approximately 79\% of the resulting data is labeled as non-gendered. During pilot experiments, we notice that it is crucial for training to utilize a balanced number of gendered and non-gendered data points. This is due the fact that if training data is unbalanced, the adversarial classifier $h$ does not effectively predict the gendered labels, and as a consequence the reverse gradients do not remove gender information. Therefore, for training adversarial networks, we create a balanced dataset by including all gendered data items and randomly sampling an equal number of the non-gendered ones. Concretely, the training process of a \modelbertadv model is as follows: we first initialize the parameters of the $f$ and $g$ components of \modelbertadv with the corresponding ones of the \modelbert ranker (\modelberttiny for \modelbertadvtiny and \modelbertmini for \modelbertadvmini), already fine-tuned on the original training data. We then freeze the parameters of $f$ and $g$, and only train the parameters of the adversarial classifier using the balanced training data. Finally, utilizing again the balanced dataset, we execute end-to-end adversarial training to jointly update all parameters.

\vspace{-0.2cm}
\paragraph{Adversarial Training Setup}
We train the \modelbertadv models for two epochs. Due to stochastic nature of adversarial training, we define 20 checkpoints during training, in which the model till that point is saved. We experiment on $\lambda$ values between $0.1$ to $0.8$ with intervals of $0.1$ for \modelbertadvtiny, and the values between $0.2$ to $0.8$ with intervals of $0.2$ for \modelbertadvmini. This results in $20\times8=160$, and and $20\times4=80$ variations for \modelbertadvtiny and \modelbertadvmini, respectively. The best result of each model according to the fairness metric is reported by conducting 5-fold cross validation.

\vspace{-0.2cm}
\paragraph{Evaluation} The fairness of the IR ranking models as well as the ranker-agnostic documents sets are evaluated with the \metricfairnorm metric with a cutoff at 10. To calculate document neutrality, we set the threshold of $\tau$ to 1 (see Eq.~\ref{eq:measurement:docfactor}). The background documents set ($\widehat{S}^q$), used to calculate \metricfairideal is set to the top 200 passages retrieved by the \modelbm model (the same as the one used by the neural models for reranking). The utility of the models are evaluated with several common metrics, namely mean reciprocal rank (\metricmrr), normalized discounted cumulative gain at cutoff 10 (\metricndcg), and \metricrecall at 10. To investigate statistical significance of results, we conduct two-sided paired $t$-tests ($p<0.05$).

\vspace{-0.1cm}
\section{Results and Analysis}
\label{sec:results}
We first focus on the evaluation results in terms of our proposed fairness metric. We then analyze the characteristics of \modelbertadv regarding fairness and utility, and report the results of a trade-off optimization approach.

\vspace{-0.2cm}
\subsection{Fairness in Retrieval Results}
\label{sec:results:fairness}
\begin{figure}[t]
\centering
\subcaptionbox{\collmsmarco}{\includegraphics[width=0.235\textwidth]{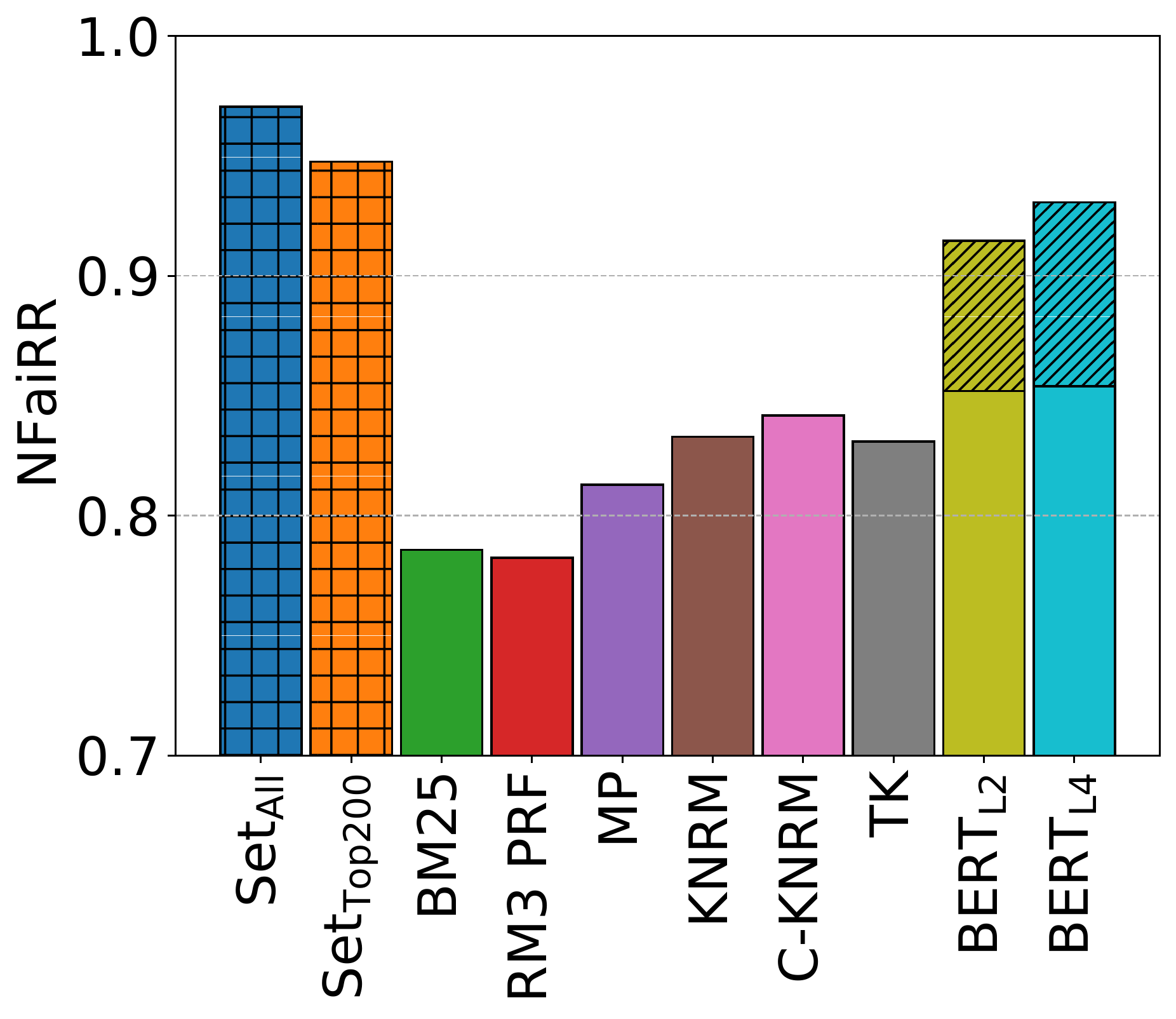}}
\subcaptionbox{\colltrecdl}{\includegraphics[width=0.235\textwidth]{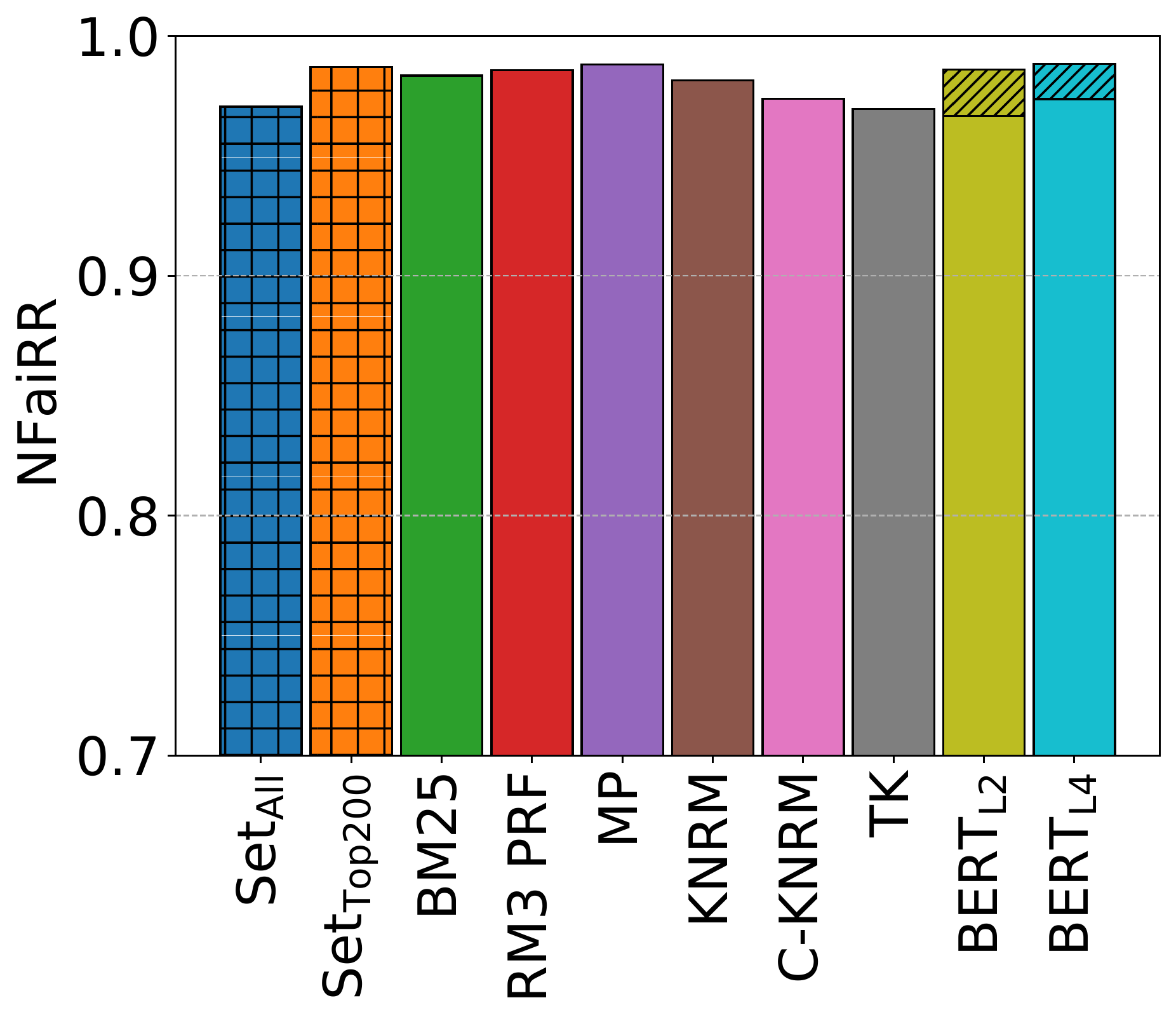}}
\centering
\vspace{-0.2cm}
\caption{Evaluation results of fairness in retrieval results. The hashed areas on top of the \modelbert rankers indicate the improvements achieved through adversarial training with the corresponding \modelbertadv models.}
\label{fig:results:fairness}
\vspace{-0.2cm}
\end{figure}

The results of the IR models as well as the two ranker-agnostic document sets according to the \metricfairnorm fairness metric are shown in Figure~\ref{fig:results:fairness}. In the plots, the fairness results of \modelberttiny and \modelbertmini are shown as the solid area of the bars, while the hashed area on top shows the improvement in fairness through the corresponding \modelbertadv models. The same results are reported in Table~\ref{tbl:results:fairness} under the \runorig columns. In the table, significant improvements of the models in \metricfairnorm in comparison with the \modelbm models are indicated with $\ddagger$, and significant improvements of \modelbertadvtiny over \modelberttiny, and \modelbertadvmini over \modelbertmini are shown with the $\dagger$ sign. 

Investigating first the ranker-agnostic document sets, we observe that in the \collmsmarco collection the \metricfairnorm of \modelrndbm is slightly lower than \modelrndcoll; in other words the ranker-agnostic fairness results of the top retrieved documents are more biased in comparison with the whole documents in the collection. This indicates that the \collmsmarco queries tend to retrieve documents from some subspaces of the collection, which contain higher gender biases in comparison with the average bias of the collection. In contrast, this is the other way around in \colltrecdl, such that \modelrndbm is more fair than \modelrndcoll and almost reaching the maximum (or ideal) fairness value. This indicates that the gender fairness-sensitive queries of the \colltrecdlorig task lead to either (almost) balanced or no representation of genders in retrieval results.

Looking at the results of the ranking models in \collmsmarco, we observe considerable differences across the IR models, while all show lower fairness scores when compared to \modelrndbm. In particular, the classical IR models (\modelbm and \modelprf) show the lowest fairness, whereas \metricfairnorm is significantly higher for all neural models. We root the cause of these differences in the more noisy results of the two classical models, which on these specific queries, result in higher degrees of gender bias. Among the neural models, the \modelbert rankers show the overall best fairness results. As reported in Table~\ref{tbl:results:fairness}, the \metricfairnorm results of both \modelbert models significantly improve in the \modelbertadv models, by a considerable margin ($7\%$ and $9\%$ in \modelbertadvtiny and \modelbertadvmini respectively), showing the effectiveness of the adversarial training method in providing more balanced rankings. Considering the results of \colltrecdl, we observe much less nuances as the starting document set \modelrndbm already provides almost fair results. Despite this, we still observe a marginal improvement in \metricfairnorm when applying our proposed adversarial training method, such that the \modelbertadvmini model offers the best fairness results in both collections.  

Let us now investigate the changes in \metricfairnorm by reordering the ranked lists with relevance judgements, as shown in the \runwqrels columns of Table~\ref{tbl:results:fairness}. The results show both increase and decrease in fairness scores, where \modelbertadv models are the only consistent ones (decreasing). Considering these results, we can not conclude any particular pattern regarding the relation between fairness and oracle rankings in respect to utility. This is, however, a fairly expected behavior, as the topic of fairness is not considered during the process of creating relevance judgements in these collections.  

For the sake of completeness, Table~\ref{tbl:results:performance} reports the evaluation results of the models according to the utility metrics, in which the significant improvements in comparison with the \modelbm models are indicated with $\ddagger$. In the following section, we discuss in detail the effect of fairness and utility in \modelbertadv models.

\begin{table}[t]
\begin{center}
\caption{\metricfairnorm results on \runorig and the changes after applying the \runwqrels setting. $\ddagger$ indicates significant improvement over \modelbm; $\dagger$ indicates significant improvement of the \modelbertadv models over their corresponding \modelbert models.}
\vspace{-0.1cm}
\centering
\begin{tabular}{ l l l l l}
\toprule
\multirow{2}{*}{Model} & \multicolumn{2}{c}{\textbf{\collmsmarco}} & \multicolumn{2}{c}{\textbf{\colltrecdl}} \\
 & \runorig & \runwqrels & \runorig & \runwqrels \\\midrule

\modelrndcoll & $0.971$ & \multicolumn{1}{c}{-} & $0.971$ & \multicolumn{1}{c}{-} \\
\modelrndbm & $0.948$ & \multicolumn{1}{c}{-}  & $0.987$ & \multicolumn{1}{c}{-} \\\midrule

\modelbm & $0.786$ & $+0.023$ & $0.984$ & $-0.019$ \\
\modelprf & $0.782$ & $+0.024$ & $0.986$ & $-0.021$ \\
\modelmp & $0.813\ddagger$ & $+0.013$ & $\textbf{0.988}$ & $-0.024$ \\
\modelknrm & $0.833\ddagger$ & $+0.011$ & $0.981$ & $-0.017$ \\
\modelconvknrm & $0.842\ddagger$ & $+0.010$ & $0.974$ & $-0.009$ \\
\modeltk & $0.831\ddagger$ & $+0.011$ & $0.970$ & $-0.005$ \\
\modelberttiny & $0.852\ddagger$ & $+0.006$ & $0.967$ & $-0.002$ \\
\modelbertadvtiny & $0.915\!\ddagger\!\dagger$ & $-0.008$ & $0.986$ & $-0.021$ \\
\modelbertmini & $0.854\ddagger$ & $+0.006$ & $0.974$ & $-0.009$ \\
\modelbertadvmini & $\textbf{0.931}\!\ddagger\!\dagger$ & $-0.014$ & $\textbf{0.988}$ & $-0.024$ \\

\bottomrule
\label{tbl:results:fairness}
\end{tabular}
\end{center}
\vspace{-0.5cm}
\end{table}

\begin{table}[t]
\begin{center}
\small
\caption{Utility evaluation of the ranking models. $\ddagger$ indicates significant improvement over \modelbm.}
\vspace{-0.2cm}
\centering
\begin{tabular}{ l c c c | c c c}
\toprule
\multirow{2}{*}{Model} & \multicolumn{3}{c}{\textbf{\collmsmarco}} & \multicolumn{3}{c}{\textbf{\colltrecdl}} \\ 
& \multicolumn{1}{c}{\metricmrr} & \multicolumn{1}{c}{\metricndcg} & \multicolumn{1}{c|}{\metricrecall} & \multicolumn{1}{c}{\metricmrr} & \multicolumn{1}{c}{\metricndcg} & \multicolumn{1}{c}{\metricrecall} \\\midrule

\modelbm & $0.107$ & $0.125$ & $0.230$ & $0.850$ & $0.534$ & $0.133$ \\
\modelprf & $0.085$ & $0.104$ & $0.209$ & $0.841$ & $0.556$ & $0.141$ \\
\modelmp & $0.169\ddagger$ & $0.191\ddagger$ & $0.297\ddagger$ & $0.961$ & $0.578$ & $0.136$ \\
\modelknrm & $0.141\ddagger$ & $0.167\ddagger$ & $0.295\ddagger$ & $0.849$ & $0.552$ & $0.137$ \\
\modelconvknrm & $0.170\ddagger$ & $0.197\ddagger$ & $0.325\ddagger$ & $0.877$ & $0.595$ & $0.144$ \\
\modeltk & $0.212\ddagger$ & $0.231\ddagger$ & $0.360\ddagger$ & $0.903$ & $0.679\ddagger$ & $0.149$ \\
\modelberttiny & $0.188\ddagger$ & $0.211\ddagger$ & $0.338\ddagger$ & $\textbf{0.939}$ & $\textbf{0.684}\ddagger$ & $0.150$ \\
\modelbertadvtiny & $0.149\ddagger$ & $0.173\ddagger$ & $0.301\ddagger$ & $0.917$ & $0.645\ddagger$ & $0.144$ \\
\modelbertmini & $\textbf{0.216}\ddagger$ & $\textbf{0.252}\ddagger$ & $\textbf{0.406}\ddagger$ & $0.933$ & $0.672\ddagger$ & $\textbf{0.154}$ \\
\modelbertadvmini & $0.160\ddagger$ & $0.197\ddagger$ & $0.360\ddagger$ & $0.903$ & $0.636\ddagger$ & $0.140$ \\

\bottomrule
\end{tabular}
\label{tbl:results:performance}
\end{center}
\vspace{-0.5cm}
\end{table}

\begin{figure}[t]
\centering
\includegraphics[width=0.3\textwidth]{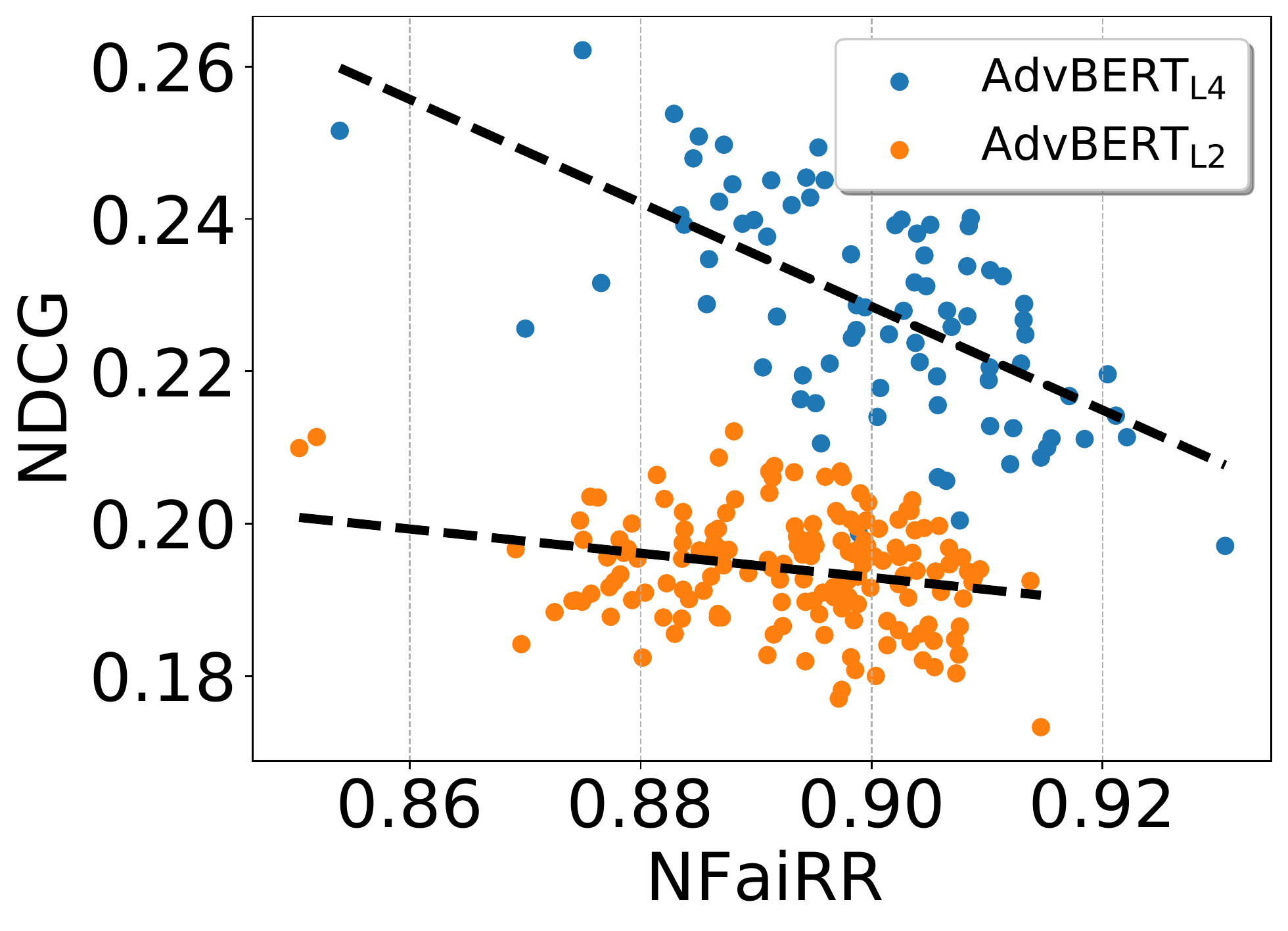}
\centering
\vspace{-0.1cm}
\caption{Fairness versus utility results on \collmsmarco for the model variations of \modelbertadv.}
\label{fig:results:advtrain}
\vspace{-0.1cm}
\end{figure}

\vspace{-0.2cm}
\subsection{Fairness -- Utility Trade-off}
The reported fairness results of the \modelbertadv models so far only consider the models with the best \metricfairnorm scores, selected through cross validation over all variations of \modelbertadv models (see Section~\ref{sec:setup} -- \emph{Adversarial Training Setup}). Despite the fact that the adversarial training method jointly optimizes both fairness and utility objectives, as reported in Table~\ref{tbl:results:performance}, the improvement in fairness is achieved with the cost of a relative decrease in the utility metrics. 

Figure~\ref{fig:results:advtrain} shows the \metricfairnorm and \metricndcg scores of all model variations of \modelbertadvtiny and \modelbertadvmini on \collmsmarco, where each point on the plot corresponds to one variation of each model.\footnote{In this section, we only report the results of \collmsmarco due to the lack of space and since this collection provides a more challenging task for studying fairness.} The dashed lines show a linear regression fitted to the scores of all variations of a model, which suggests the overall trend: As \metricfairnorm increases, \metricndcg generally decreases; and this decrease is more steep for \modelbertadvmini than for \modelbertadvtiny.

\vspace{-0.1cm}
\subsubsection{A Combinatorial Metric for Model Selection} In practice, it is crucial to follow a principled approach for model selection when it comes to a trade-off between fairness and utility. To this end, we first define a combinatorial metric based on the formulation of the well-known $F_{\beta}$ score, to combine the fairness metric (\metricfairnorm) with a utility one (\metricndcg), as formulated in the following:
\begin{equation}
F_{\beta} = (1+\beta^2)\frac{\Delta{\metricndcg} \cdot \Delta{\metricfairnorm}}{(\beta^2 \cdot \Delta{\metricndcg}) + \Delta{\metricfairnorm}}
\label{eq:measurement:metricfbeta}
\end{equation}
where $\Delta{\metricndcg}$ for a variation of a model is the difference of the \metricndcg of the variation from the minimum \metricndcg score among all variations of the model, scaled by min-max normalization (the same is applied to $\Delta{\metricfairnorm}$). The $\beta$ hyper-parameter acts as a leverage (in the hand of practitioners) to impose the preference in the model selection process, inclined towards fairness or utility. Specifically, $\beta=0$ indicates full preference towards utility, where the selected \modelbertadv model most probably becomes the same as \modelbert (no bias mitigation). $\beta=1$ gives equal importance to fairness and utility. Higher $\beta$ values give proportionally more importance to fairness than to utility. Setting $\beta$ to $\infty$ (or in fact in practice to a very large number) results in the selection of the variation of the \modelbertadvtiny or \modelbertadvmini model with the highest \metricfairnorm score, which are equivalent to the ones reported in Section~\ref{sec:results:fairness}.  

Upon deciding on a $\beta$ value, the standard model selection process is conducted: the $F_{\beta}$ score is calculated for all the variations of a given model, and the model variation with the highest $F_{\beta}$ score is selected. In the following experiments, we report the corresponding fairness and utility scores achieved through this model selection by applying 5-fold cross validation on the \collmsmarco queries.

\vspace{-0.1cm}
\subsubsection{Final Results} Table~\ref{tbl:results:fairnessperformance} shows the evaluation results of the fairness and utility metrics of \modelbertadvtiny and \modelbertadvmini for a range of $\beta$ values from $0.0$ (no bias mitigation -- highest utility) to $\infty$ (maximum fairness). In the table, the results of $\beta=0.0$ are in fact the same as the ones of the \modelbert models, which we consider as baselines. For each adversarial model, The $\dagger$ sign on \metricfairnorm indicates a significant increase in fairness, while on utility metrics (\metricmrr, \metricndcg, \metricrecall) it shows a significant decrease in scores, when compared with the results of the corresponding \modelbert model.

As shown, in both models the highest fairness scores result in a significant loss in utility metrics. However, for several lower $\beta$ values, we observe significant improvements in \metricfairnorm with no significant loss in utility metrics. In particular, the proper ranges of $\beta$ that satisfy both fairness and utility are indicated in Table~\ref{tbl:results:fairnessperformance} between the dashed lines for each model. These results highlight the effectiveness of our adversarial training approach, which -- when combined with the discussed model selection method -- provide significantly more fair models without significant loss in utility.

\begin{table}[t]
\begin{center}
\vspace{-0.2cm}
\caption{Changes in fairness and utility metrics with different values of $\beta$. The significant improvements of \metricfairnorm and the significant loss in utility metrics in comparison with the \modelbert models are indicated with the $\dagger$ sign.}
\vspace{-0.2cm}
\centering
\begin{tabular}{ l c l | l l l }
\toprule
\multicolumn{1}{c}{Model} & $\beta$ &  \multicolumn{1}{c}{\metricfairnorm} & \multicolumn{1}{c}{\metricmrr} & \multicolumn{1}{c}{\metricndcg} & \multicolumn{1}{c}{\metricrecall} \\\midrule

\multicolumn{1}{r}{\modelberttiny$\rightarrow$} & 0.0 & $0.850$ & $0.188$ & $0.211$ & $0.338$ \\\cdashlinelr{2-6}
\multirow{5}{*}{\modelbertadvtiny} & 0.2 & $0.888\dagger$ & $0.180$ & $0.212$ & $0.364$ \\
 & 0.5 & $0.888\dagger$ & $0.180$ & $0.212$ & $0.364$ \\
 & 1.0 & $0.904\dagger$ & $0.167$ & $0.203$ & $0.357$ \\
 & 2.0 & $0.914\dagger$ & $0.171$ & $0.193$ & $0.319$ \\
 & 5.0 & $0.914\dagger$ & $0.171$ & $0.193$ & $0.319$ \\\cdashlinelr{2-6}
 & $\infty$ & $0.915\dagger$ & $0.149\dagger$ & $0.173\dagger$ & $0.301$ \\\midrule

\multicolumn{1}{r}{\modelbertmini$\rightarrow$} & 0.0 & $0.854$ & $0.216$ & $0.252$ & $0.406$ \\\cdashlinelr{2-6}
\multirow{5}{*}{\modelbertadvmini} & 0.2 & $0.900\dagger$ & $0.215$ & $0.258$ & $0.434$ \\
 & 0.5 & $0.900\dagger$ & $0.215$ & $0.258$ & $0.434$ \\
 & 1.0 & $0.900\dagger$ & $0.215$ & $0.258$ & $0.434$ \\
 & 2.0 & $0.909\dagger$ & $0.214$ & $0.240$ & $0.369$ \\\cdashlinelr{2-6}
 & 5.0 & $0.920\dagger$ & $0.184\dagger$ & $0.220\dagger$ & $0.376$ \\
 & $\infty$ & $0.931\dagger$ & $0.160\dagger$ & $0.197\dagger$ & $0.360$ \\

\bottomrule
\label{tbl:results:fairnessperformance}
\end{tabular}
\vspace{-1.2cm}
\end{center}
\end{table}



\vspace{-0.2cm}
\section{Conclusion and Outlook}\label{sec:conclusions}
This work provides a standard benchmark for measuring fairness in retrieval results, and proposes an adversarial training method to mitigate bias in \modelbert ranking models. The benchmark puts forward a generic framework, consisting of metrics for measuring fairness in a given ranked list as well as in a subset of collection's documents. Through human annotation, we provide fairness-sensitive subsets of the queries of two recent passage retrieval collections, \collmsmarco and \colltrecdl, enabling reproducible research on fairness of IR models together with their utilities. Our experimental results show that, in the more fairness-challenging \collmsmarco collection, the results of IR rankers are more gender-biased in comparison with the ranker-agnostic baselines, while the fairness of \modelbert rankers significantly improves by applying our proposed adversarial training. Finally, through a principled model selection method, we show that the resulting \modelbertadv models can effectively maintain the significant improvements in fairness with no significant loss in the utility metrics. Future research will investigate the relations between the introduced fairness metrics and the human perception of bias and fairness in retrieved contents, as well as other algorithmic bias mitigation approaches.

\section*{Acknowledgements}
Thanks to Klara Krieg for her help on creating the dataset. This work was funded by the Know-Center GmbH (FFG COMET program) and the H2020 projects TRIPLE (GA: 863420) and AI4EU (GA: 825619).

\bibliographystyle{ACM-Reference-Format}
\balance
\bibliography{references}


\begin{thebibliography}{61}


\ifx \showCODEN    \undefined \def \showCODEN     #1{\unskip}     \fi
\ifx \showDOI      \undefined \def \showDOI       #1{#1}\fi
\ifx \showISBNx    \undefined \def \showISBNx     #1{\unskip}     \fi
\ifx \showISBNxiii \undefined \def \showISBNxiii  #1{\unskip}     \fi
\ifx \showISSN     \undefined \def \showISSN      #1{\unskip}     \fi
\ifx \showLCCN     \undefined \def \showLCCN      #1{\unskip}     \fi
\ifx \shownote     \undefined \def \shownote      #1{#1}          \fi
\ifx \showarticletitle \undefined \def \showarticletitle #1{#1}   \fi
\ifx \showURL      \undefined \def \showURL       {\relax}        \fi
\providecommand\bibfield[2]{#2}
\providecommand\bibinfo[2]{#2}
\providecommand\natexlab[1]{#1}
\providecommand\showeprint[2][]{arXiv:#2}

\bibitem[\protect\citeauthoryear{Abdul-Jaleel, Allan, Croft, Diaz, Larkey, Li,
  Smucker, and Wade}{Abdul-Jaleel et~al\mbox{.}}{2004}]%
        {abdul2004umass}
\bibfield{author}{\bibinfo{person}{Nasreen Abdul-Jaleel},
  \bibinfo{person}{James Allan}, \bibinfo{person}{W~Bruce Croft},
  \bibinfo{person}{Fernando Diaz}, \bibinfo{person}{Leah Larkey},
  \bibinfo{person}{Xiaoyan Li}, \bibinfo{person}{Mark~D Smucker}, {and}
  \bibinfo{person}{Courtney Wade}.} \bibinfo{year}{2004}\natexlab{}.
\newblock \showarticletitle{UMass at TREC 2004: Novelty and HARD}.
\newblock \bibinfo{journal}{\emph{Computer Science Department Faculty
  Publication Series}} (\bibinfo{year}{2004}), \bibinfo{pages}{189}.
\newblock


\bibitem[\protect\citeauthoryear{Baeza-Yates}{Baeza-Yates}{2018}]%
        {baeza2018bias}
\bibfield{author}{\bibinfo{person}{Ricardo Baeza-Yates}.}
  \bibinfo{year}{2018}\natexlab{}.
\newblock \showarticletitle{Bias on the web}.
\newblock \bibinfo{journal}{\emph{Commun. ACM}} (\bibinfo{year}{2018}).
\newblock


\bibitem[\protect\citeauthoryear{Baeza-Yates}{Baeza-Yates}{2020}]%
        {baeza2020bias}
\bibfield{author}{\bibinfo{person}{Ricardo Baeza-Yates}.}
  \bibinfo{year}{2020}\natexlab{}.
\newblock \showarticletitle{Bias in search and recommender systems}. In
  \bibinfo{booktitle}{\emph{Proceedings of the ACM Conference on Recommender
  Systems}}. \bibinfo{pages}{2--2}.
\newblock


\bibitem[\protect\citeauthoryear{Barrett, Kementchedjhieva, Elazar, Elliott,
  and S{\o}gaard}{Barrett et~al\mbox{.}}{2019}]%
        {barrett2019adversarial}
\bibfield{author}{\bibinfo{person}{Maria Barrett}, \bibinfo{person}{Yova
  Kementchedjhieva}, \bibinfo{person}{Yanai Elazar}, \bibinfo{person}{Desmond
  Elliott}, {and} \bibinfo{person}{Anders S{\o}gaard}.}
  \bibinfo{year}{2019}\natexlab{}.
\newblock \showarticletitle{Adversarial removal of demographic attributes
  revisited}. In \bibinfo{booktitle}{\emph{Proceedings of the Conference on
  Empirical Methods in Natural Language Processing and the 9th International
  Joint Conference on Natural Language Processing (EMNLP-IJCNLP)}}.
  \bibinfo{pages}{6331--6336}.
\newblock


\bibitem[\protect\citeauthoryear{Biega, Gummadi, and Weikum}{Biega
  et~al\mbox{.}}{2018}]%
        {biega2018equity}
\bibfield{author}{\bibinfo{person}{Asia~J Biega}, \bibinfo{person}{Krishna~P
  Gummadi}, {and} \bibinfo{person}{Gerhard Weikum}.}
  \bibinfo{year}{2018}\natexlab{}.
\newblock \showarticletitle{Equity of attention: Amortizing individual fairness
  in rankings}. In \bibinfo{booktitle}{\emph{The 41st international ACM SIGIR
  Conference on Research \& Development in Information Retrieval}}.
  \bibinfo{pages}{405--414}.
\newblock


\bibitem[\protect\citeauthoryear{Blodgett, Barocas, Daum{\'e}~III, and
  Wallach}{Blodgett et~al\mbox{.}}{2020}]%
        {blodgett2020language}
\bibfield{author}{\bibinfo{person}{Su~Lin Blodgett}, \bibinfo{person}{Solon
  Barocas}, \bibinfo{person}{Hal Daum{\'e}~III}, {and} \bibinfo{person}{Hanna
  Wallach}.} \bibinfo{year}{2020}\natexlab{}.
\newblock \showarticletitle{Language (Technology) is Power: A Critical Survey
  of “Bias” in NLP}. In \bibinfo{booktitle}{\emph{Proceedings of the 58th
  Annual Meeting of the Association for Computational Linguistics}}.
  \bibinfo{pages}{5454--5476}.
\newblock


\bibitem[\protect\citeauthoryear{Bolukbasi, Chang, Zou, Saligrama, and
  Kalai}{Bolukbasi et~al\mbox{.}}{2016}]%
        {bolukbasi2016man}
\bibfield{author}{\bibinfo{person}{Tolga Bolukbasi}, \bibinfo{person}{Kai-Wei
  Chang}, \bibinfo{person}{James~Y Zou}, \bibinfo{person}{Venkatesh Saligrama},
  {and} \bibinfo{person}{Adam~T Kalai}.} \bibinfo{year}{2016}\natexlab{}.
\newblock \showarticletitle{Man is to computer programmer as woman is to
  homemaker? debiasing word embeddings}.
\newblock \bibinfo{journal}{\emph{Advances in Neural Information Processing
  Systems}} (\bibinfo{year}{2016}).
\newblock


\bibitem[\protect\citeauthoryear{Caliskan, Bryson, and Narayanan}{Caliskan
  et~al\mbox{.}}{2017}]%
        {caliskan2017semantics}
\bibfield{author}{\bibinfo{person}{Aylin Caliskan}, \bibinfo{person}{Joanna~J
  Bryson}, {and} \bibinfo{person}{Arvind Narayanan}.}
  \bibinfo{year}{2017}\natexlab{}.
\newblock \showarticletitle{Semantics derived automatically from language
  corpora contain human-like biases}.
\newblock \bibinfo{journal}{\emph{Science}} (\bibinfo{year}{2017}).
\newblock


\bibitem[\protect\citeauthoryear{Celis, Straszak, and Vishnoi}{Celis
  et~al\mbox{.}}{2018}]%
        {celis2018ranking}
\bibfield{author}{\bibinfo{person}{L~Elisa Celis}, \bibinfo{person}{Damian
  Straszak}, {and} \bibinfo{person}{Nisheeth~K Vishnoi}.}
  \bibinfo{year}{2018}\natexlab{}.
\newblock \showarticletitle{Ranking with Fairness Constraints}. In
  \bibinfo{booktitle}{\emph{Proceedings of the 45th International Colloquium on
  Automata, Languages, and Programming (ICALP)}}. Schloss
  Dagstuhl-Leibniz-Zentrum fuer Informatik.
\newblock


\bibitem[\protect\citeauthoryear{Chen, Ma, Hann{\'a}k, and Wilson}{Chen
  et~al\mbox{.}}{2018}]%
        {chen2018investigating}
\bibfield{author}{\bibinfo{person}{Le Chen}, \bibinfo{person}{Ruijun Ma},
  \bibinfo{person}{Anik{\'o} Hann{\'a}k}, {and} \bibinfo{person}{Christo
  Wilson}.} \bibinfo{year}{2018}\natexlab{}.
\newblock \showarticletitle{Investigating the impact of gender on rank in
  resume search engines}. In \bibinfo{booktitle}{\emph{Proceedings of the 2018
  CHI Conference on Human Factors in Computing Systems}}.
  \bibinfo{pages}{1--14}.
\newblock


\bibitem[\protect\citeauthoryear{Craswell, Mitra, Yilmaz, Campos, and
  Voorhees}{Craswell et~al\mbox{.}}{2020}]%
        {craswell2020overview}
\bibfield{author}{\bibinfo{person}{Nick Craswell}, \bibinfo{person}{Bhaskar
  Mitra}, \bibinfo{person}{Emine Yilmaz}, \bibinfo{person}{Daniel Campos},
  {and} \bibinfo{person}{Ellen~M Voorhees}.} \bibinfo{year}{2020}\natexlab{}.
\newblock \showarticletitle{Overview of the trec 2019 deep learning track}.
\newblock \bibinfo{journal}{\emph{arXiv preprint arXiv:2003.07820}}
  (\bibinfo{year}{2020}).
\newblock


\bibitem[\protect\citeauthoryear{Dai, Xiong, Callan, and Liu}{Dai
  et~al\mbox{.}}{2018}]%
        {dai2018convolutional}
\bibfield{author}{\bibinfo{person}{Zhuyun Dai}, \bibinfo{person}{Chenyan
  Xiong}, \bibinfo{person}{Jamie Callan}, {and} \bibinfo{person}{Zhiyuan Liu}.}
  \bibinfo{year}{2018}\natexlab{}.
\newblock \showarticletitle{Convolutional neural networks for soft-matching
  n-grams in ad-hoc search}. In \bibinfo{booktitle}{\emph{Proceedings of the
  11th ACM International Conference on Web Search and Data Mining}}.
  \bibinfo{pages}{126--134}.
\newblock


\bibitem[\protect\citeauthoryear{De-Arteaga, Romanov, Wallach, Chayes, Borgs,
  Chouldechova, Geyik, Kenthapadi, and Kalai}{De-Arteaga et~al\mbox{.}}{2019}]%
        {de2019bias}
\bibfield{author}{\bibinfo{person}{Maria De-Arteaga}, \bibinfo{person}{Alexey
  Romanov}, \bibinfo{person}{Hanna Wallach}, \bibinfo{person}{Jennifer Chayes},
  \bibinfo{person}{Christian Borgs}, \bibinfo{person}{Alexandra Chouldechova},
  \bibinfo{person}{Sahin Geyik}, \bibinfo{person}{Krishnaram Kenthapadi}, {and}
  \bibinfo{person}{Adam~Tauman Kalai}.} \bibinfo{year}{2019}\natexlab{}.
\newblock \showarticletitle{Bias in bios: A case study of semantic
  representation bias in a high-stakes setting}. In
  \bibinfo{booktitle}{\emph{Proceedings of the Conference on Fairness,
  Accountability, and Transparency}}. \bibinfo{pages}{120--128}.
\newblock


\bibitem[\protect\citeauthoryear{Devlin, Chang, Lee, and Toutanova}{Devlin
  et~al\mbox{.}}{2019}]%
        {devlin2019bert}
\bibfield{author}{\bibinfo{person}{Jacob Devlin}, \bibinfo{person}{Ming-Wei
  Chang}, \bibinfo{person}{Kenton Lee}, {and} \bibinfo{person}{Kristina
  Toutanova}.} \bibinfo{year}{2019}\natexlab{}.
\newblock \showarticletitle{BERT: Pre-training of Deep Bidirectional
  Transformers for Language Understanding}. In
  \bibinfo{booktitle}{\emph{Proceedings of the 2019 Conference of the North
  American Chapter of the Association for Computational Linguistics: Human
  Language Technologies}}. \bibinfo{pages}{4171--4186}.
\newblock


\bibitem[\protect\citeauthoryear{Ekstrand, Burke, and Diaz}{Ekstrand
  et~al\mbox{.}}{2019}]%
        {ekstrand2019fairness}
\bibfield{author}{\bibinfo{person}{Michael~D Ekstrand}, \bibinfo{person}{Robin
  Burke}, {and} \bibinfo{person}{Fernando Diaz}.}
  \bibinfo{year}{2019}\natexlab{}.
\newblock \showarticletitle{Fairness and discrimination in retrieval and
  recommendation}. In \bibinfo{booktitle}{\emph{Proceedings of the 42nd
  International ACM SIGIR Conference on Research and Development in Information
  Retrieval}}. \bibinfo{pages}{1403--1404}.
\newblock


\bibitem[\protect\citeauthoryear{Elazar and Goldberg}{Elazar and
  Goldberg}{2018}]%
        {elazar2018adversarial}
\bibfield{author}{\bibinfo{person}{Yanai Elazar} {and} \bibinfo{person}{Yoav
  Goldberg}.} \bibinfo{year}{2018}\natexlab{}.
\newblock \showarticletitle{Adversarial Removal of Demographic Attributes from
  Text Data}. In \bibinfo{booktitle}{\emph{Proceedings of the Conference on
  Empirical Methods in Natural Language Processing}}. \bibinfo{pages}{11--21}.
\newblock


\bibitem[\protect\citeauthoryear{Fabris, Purpura, Silvello, and Susto}{Fabris
  et~al\mbox{.}}{2020}]%
        {fabris2020gender}
\bibfield{author}{\bibinfo{person}{Alessandro Fabris}, \bibinfo{person}{Alberto
  Purpura}, \bibinfo{person}{Gianmaria Silvello}, {and}
  \bibinfo{person}{Gian~Antonio Susto}.} \bibinfo{year}{2020}\natexlab{}.
\newblock \showarticletitle{Gender stereotype reinforcement: Measuring the
  gender bias conveyed by ranking algorithms}.
\newblock \bibinfo{journal}{\emph{Information Processing \& Management}}
  (\bibinfo{year}{2020}).
\newblock


\bibitem[\protect\citeauthoryear{Ganin and Lempitsky}{Ganin and
  Lempitsky}{2015}]%
        {ganin2015unsupervised}
\bibfield{author}{\bibinfo{person}{Yaroslav Ganin} {and}
  \bibinfo{person}{Victor Lempitsky}.} \bibinfo{year}{2015}\natexlab{}.
\newblock \showarticletitle{Unsupervised domain adaptation by backpropagation}.
  In \bibinfo{booktitle}{\emph{Proceedings of the International Conference on
  Machine Learning}}. PMLR, \bibinfo{pages}{1180--1189}.
\newblock


\bibitem[\protect\citeauthoryear{Gao and Shah}{Gao and Shah}{2020}]%
        {gao2020toward}
\bibfield{author}{\bibinfo{person}{Ruoyuan Gao} {and} \bibinfo{person}{Chirag
  Shah}.} \bibinfo{year}{2020}\natexlab{}.
\newblock \showarticletitle{Toward creating a fairer ranking in search engine
  results}.
\newblock \bibinfo{journal}{\emph{Information Processing \& Management}}
  (\bibinfo{year}{2020}), \bibinfo{pages}{102138}.
\newblock


\bibitem[\protect\citeauthoryear{Gerritse, Hasibi, and de~Vries}{Gerritse
  et~al\mbox{.}}{2020}]%
        {gerritse2020bias}
\bibfield{author}{\bibinfo{person}{Emma~J Gerritse}, \bibinfo{person}{Faegheh
  Hasibi}, {and} \bibinfo{person}{Arjen~P de Vries}.}
  \bibinfo{year}{2020}\natexlab{}.
\newblock \showarticletitle{Bias in Conversational Search: The Double-Edged
  Sword of the Personalized Knowledge Graph}. In
  \bibinfo{booktitle}{\emph{Proceedings of the 2020 ACM SIGIR on International
  Conference on Theory of Information Retrieval}}. \bibinfo{pages}{133--136}.
\newblock


\bibitem[\protect\citeauthoryear{Geyik, Ambler, and Kenthapadi}{Geyik
  et~al\mbox{.}}{2019}]%
        {geyik2019fairness}
\bibfield{author}{\bibinfo{person}{Sahin~Cem Geyik}, \bibinfo{person}{Stuart
  Ambler}, {and} \bibinfo{person}{Krishnaram Kenthapadi}.}
  \bibinfo{year}{2019}\natexlab{}.
\newblock \showarticletitle{Fairness-aware ranking in search \& recommendation
  systems with application to linkedin talent search}. In
  \bibinfo{booktitle}{\emph{Proceedings of the 25th ACM SIGKDD International
  Conference on Knowledge Discovery \& Data Mining}}.
  \bibinfo{pages}{2221--2231}.
\newblock


\bibitem[\protect\citeauthoryear{Goodfellow, Pouget-Abadie, Mirza, Xu,
  Warde-Farley, Ozair, Courville, and Bengio}{Goodfellow et~al\mbox{.}}{2014}]%
        {goodfellow2014generative}
\bibfield{author}{\bibinfo{person}{Ian~J Goodfellow}, \bibinfo{person}{Jean
  Pouget-Abadie}, \bibinfo{person}{Mehdi Mirza}, \bibinfo{person}{Bing Xu},
  \bibinfo{person}{David Warde-Farley}, \bibinfo{person}{Sherjil Ozair},
  \bibinfo{person}{Aaron Courville}, {and} \bibinfo{person}{Yoshua Bengio}.}
  \bibinfo{year}{2014}\natexlab{}.
\newblock \showarticletitle{Generative adversarial nets}. In
  \bibinfo{booktitle}{\emph{Proceedings of the 27th International Conference on
  Neural Information Processing Systems-Volume 2}}.
  \bibinfo{pages}{2672--2680}.
\newblock


\bibitem[\protect\citeauthoryear{Headquarters}{Headquarters}{2020}]%
        {UNSDG}
\bibfield{author}{\bibinfo{person}{UN~Women Headquarters}.}
  \bibinfo{year}{2020}\natexlab{}.
\newblock \bibinfo{title}{{UN Women} Gender equality: Women's rights in review
  25 years after Beijing}.
\newblock
  \bibinfo{howpublished}{\url{https://www.unwomen.org/en/digital-library/publications/2020/03/womens-rights-in-review}}.
\newblock
\newblock
\shownote{Accessed: 2021-02-06.}


\bibitem[\protect\citeauthoryear{Hofst{\"a}tter, Rekabsaz, Eickhoff, and
  Hanbury}{Hofst{\"a}tter et~al\mbox{.}}{2019}]%
        {hofstatter2019effect}
\bibfield{author}{\bibinfo{person}{Sebastian Hofst{\"a}tter},
  \bibinfo{person}{Navid Rekabsaz}, \bibinfo{person}{Carsten Eickhoff}, {and}
  \bibinfo{person}{Allan Hanbury}.} \bibinfo{year}{2019}\natexlab{}.
\newblock \showarticletitle{On the effect of low-frequency terms on neural-IR
  models}. In \bibinfo{booktitle}{\emph{Proceedings of the 42nd International
  ACM SIGIR Conference on Research and Development in Information Retrieval}}.
  \bibinfo{pages}{1137--1140}.
\newblock


\bibitem[\protect\citeauthoryear{Hofst{\"a}tter, Zamani, Mitra, Craswell, and
  Hanbury}{Hofst{\"a}tter et~al\mbox{.}}{2020}]%
        {Hofstaetter2020_sigir}
\bibfield{author}{\bibinfo{person}{Sebastian Hofst{\"a}tter},
  \bibinfo{person}{Hamed Zamani}, \bibinfo{person}{Bhaskar Mitra},
  \bibinfo{person}{Nick Craswell}, {and} \bibinfo{person}{Allan Hanbury}.}
  \bibinfo{year}{2020}\natexlab{}.
\newblock \showarticletitle{{Local Self-Attention over Long Text for Efficient
  Document Retrieval}}. In \bibinfo{booktitle}{\emph{Proceedings of the ACM
  SIGIR Conference on Research and Development in Information Retrieval}}.
\newblock


\bibitem[\protect\citeauthoryear{J{\"a}rvelin and
  Kek{\"a}l{\"a}inen}{J{\"a}rvelin and Kek{\"a}l{\"a}inen}{2002}]%
        {jarvelin2002cumulated}
\bibfield{author}{\bibinfo{person}{Kalervo J{\"a}rvelin} {and}
  \bibinfo{person}{Jaana Kek{\"a}l{\"a}inen}.} \bibinfo{year}{2002}\natexlab{}.
\newblock \showarticletitle{Cumulated gain-based evaluation of IR techniques}.
\newblock \bibinfo{journal}{\emph{ACM Transactions on Information Systems
  (TOIS)}} (\bibinfo{year}{2002}), \bibinfo{pages}{422--446}.
\newblock


\bibitem[\protect\citeauthoryear{Kay, Matuszek, and Munson}{Kay
  et~al\mbox{.}}{2015}]%
        {kay2015unequal}
\bibfield{author}{\bibinfo{person}{Matthew Kay}, \bibinfo{person}{Cynthia
  Matuszek}, {and} \bibinfo{person}{Sean~A Munson}.}
  \bibinfo{year}{2015}\natexlab{}.
\newblock \showarticletitle{Unequal representation and gender stereotypes in
  image search results for occupations}. In
  \bibinfo{booktitle}{\emph{Proceedings of the 33rd Annual ACM Conference on
  Human Factors in Computing Systems}}. \bibinfo{pages}{3819--3828}.
\newblock


\bibitem[\protect\citeauthoryear{Khattab and Zaharia}{Khattab and
  Zaharia}{2020}]%
        {khattab2020colbert}
\bibfield{author}{\bibinfo{person}{Omar Khattab} {and} \bibinfo{person}{Matei
  Zaharia}.} \bibinfo{year}{2020}\natexlab{}.
\newblock \showarticletitle{Colbert: Efficient and effective passage search via
  contextualized late interaction over bert}. In
  \bibinfo{booktitle}{\emph{Proceedings of the 43rd International ACM SIGIR
  Conference on Research and Development in Information Retrieval}}.
  \bibinfo{pages}{39--48}.
\newblock


\bibitem[\protect\citeauthoryear{Kulshrestha, Eslami, Messias, Zafar, Ghosh,
  Gummadi, and Karahalios}{Kulshrestha et~al\mbox{.}}{2017}]%
        {kulshrestha2017quantifying}
\bibfield{author}{\bibinfo{person}{Juhi Kulshrestha},
  \bibinfo{person}{Motahhare Eslami}, \bibinfo{person}{Johnnatan Messias},
  \bibinfo{person}{Muhammad~Bilal Zafar}, \bibinfo{person}{Saptarshi Ghosh},
  \bibinfo{person}{Krishna~P Gummadi}, {and} \bibinfo{person}{Karrie
  Karahalios}.} \bibinfo{year}{2017}\natexlab{}.
\newblock \showarticletitle{Quantifying search bias: Investigating sources of
  bias for political searches in social media}. In
  \bibinfo{booktitle}{\emph{Proceedings of the 2017 ACM Conference on Computer
  Supported Cooperative Work and Social Computing}}. \bibinfo{pages}{417--432}.
\newblock


\bibitem[\protect\citeauthoryear{Lorber}{Lorber}{2005}]%
        {lorber2005breaking}
\bibfield{author}{\bibinfo{person}{Judith Lorber}.}
  \bibinfo{year}{2005}\natexlab{}.
\newblock \bibinfo{booktitle}{\emph{Breaking the bowls: Degendering and
  feminist change}}.
\newblock


\bibitem[\protect\citeauthoryear{MacAvaney, Nardini, Perego, Tonellotto,
  Goharian, and Frieder}{MacAvaney et~al\mbox{.}}{2020}]%
        {macavaney2020efficient}
\bibfield{author}{\bibinfo{person}{Sean MacAvaney},
  \bibinfo{person}{Franco~Maria Nardini}, \bibinfo{person}{Raffaele Perego},
  \bibinfo{person}{Nicola Tonellotto}, \bibinfo{person}{Nazli Goharian}, {and}
  \bibinfo{person}{Ophir Frieder}.} \bibinfo{year}{2020}\natexlab{}.
\newblock \showarticletitle{Efficient document re-ranking for transformers by
  precomputing term representations}. In \bibinfo{booktitle}{\emph{Proceedings
  of the 43rd International ACM SIGIR Conference on Research and Development in
  Information Retrieval}}. \bibinfo{pages}{49--58}.
\newblock


\bibitem[\protect\citeauthoryear{Madras, Creager, Pitassi, and Zemel}{Madras
  et~al\mbox{.}}{2018}]%
        {madras2018learning}
\bibfield{author}{\bibinfo{person}{David Madras}, \bibinfo{person}{Elliot
  Creager}, \bibinfo{person}{Toniann Pitassi}, {and} \bibinfo{person}{Richard
  Zemel}.} \bibinfo{year}{2018}\natexlab{}.
\newblock \showarticletitle{Learning adversarially fair and transferable
  representations}. In \bibinfo{booktitle}{\emph{Proceedings of the
  International Conference on Machine Learning}}. PMLR,
  \bibinfo{pages}{3384--3393}.
\newblock


\bibitem[\protect\citeauthoryear{Maudslay, Gonen, Cotterell, and
  Teufel}{Maudslay et~al\mbox{.}}{2019}]%
        {maudslay2019s}
\bibfield{author}{\bibinfo{person}{Rowan~Hall Maudslay}, \bibinfo{person}{Hila
  Gonen}, \bibinfo{person}{Ryan Cotterell}, {and} \bibinfo{person}{Simone
  Teufel}.} \bibinfo{year}{2019}\natexlab{}.
\newblock \showarticletitle{It’s All in the Name: Mitigating Gender Bias with
  Name-Based Counterfactual Data Substitution}. In
  \bibinfo{booktitle}{\emph{Proceedings of the Conference on Empirical Methods
  in Natural Language Processing and the 9th International Joint Conference on
  Natural Language Processing (EMNLP-IJCNLP)}}. \bibinfo{pages}{5270--5278}.
\newblock


\bibitem[\protect\citeauthoryear{Mehrabi, Morstatter, Saxena, Lerman, and
  Galstyan}{Mehrabi et~al\mbox{.}}{2019}]%
        {mehrabi2019survey}
\bibfield{author}{\bibinfo{person}{Ninareh Mehrabi}, \bibinfo{person}{Fred
  Morstatter}, \bibinfo{person}{Nripsuta Saxena}, \bibinfo{person}{Kristina
  Lerman}, {and} \bibinfo{person}{Aram Galstyan}.}
  \bibinfo{year}{2019}\natexlab{}.
\newblock \showarticletitle{A survey on bias and fairness in machine learning}.
\newblock \bibinfo{journal}{\emph{arXiv preprint arXiv:1908.09635}}
  (\bibinfo{year}{2019}).
\newblock


\bibitem[\protect\citeauthoryear{Morik, Singh, Hong, and Joachims}{Morik
  et~al\mbox{.}}{2020}]%
        {morik2020controlling}
\bibfield{author}{\bibinfo{person}{Marco Morik}, \bibinfo{person}{Ashudeep
  Singh}, \bibinfo{person}{Jessica Hong}, {and} \bibinfo{person}{Thorsten
  Joachims}.} \bibinfo{year}{2020}\natexlab{}.
\newblock \showarticletitle{Controlling fairness and bias in dynamic
  learning-to-rank}. In \bibinfo{booktitle}{\emph{Proceedings of the 43rd
  International ACM SIGIR Conference on Research and Development in Information
  Retrieval}}. \bibinfo{pages}{429--438}.
\newblock


\bibitem[\protect\citeauthoryear{Nguyen, Rosenberg, Song, Gao, Tiwary,
  Majumder, and Deng}{Nguyen et~al\mbox{.}}{2016}]%
        {nguyen2016ms}
\bibfield{author}{\bibinfo{person}{Tri Nguyen}, \bibinfo{person}{Mir
  Rosenberg}, \bibinfo{person}{Xia Song}, \bibinfo{person}{Jianfeng Gao},
  \bibinfo{person}{Saurabh Tiwary}, \bibinfo{person}{Rangan Majumder}, {and}
  \bibinfo{person}{Li Deng}.} \bibinfo{year}{2016}\natexlab{}.
\newblock \showarticletitle{{MS MARCO}: A human generated machine reading
  comprehension dataset}.
\newblock \bibinfo{journal}{\emph{arXiv preprint arXiv:1611.09268}}
  (\bibinfo{year}{2016}).
\newblock


\bibitem[\protect\citeauthoryear{Nogueira and Cho}{Nogueira and Cho}{2019}]%
        {nogueira2019passage}
\bibfield{author}{\bibinfo{person}{Rodrigo Nogueira} {and}
  \bibinfo{person}{Kyunghyun Cho}.} \bibinfo{year}{2019}\natexlab{}.
\newblock \showarticletitle{Passage Re-ranking with BERT}.
\newblock \bibinfo{journal}{\emph{arXiv preprint arXiv:1901.04085}}
  (\bibinfo{year}{2019}).
\newblock


\bibitem[\protect\citeauthoryear{Otterbacher, Bates, and Clough}{Otterbacher
  et~al\mbox{.}}{2017}]%
        {otterbacher2017competent}
\bibfield{author}{\bibinfo{person}{Jahna Otterbacher}, \bibinfo{person}{Jo
  Bates}, {and} \bibinfo{person}{Paul Clough}.}
  \bibinfo{year}{2017}\natexlab{}.
\newblock \showarticletitle{Competent men and warm women: Gender stereotypes
  and backlash in image search results}. In
  \bibinfo{booktitle}{\emph{Proceedings of the 2017 CHI Conference on Human
  Factors in Computing Systems}}. \bibinfo{pages}{6620--6631}.
\newblock


\bibitem[\protect\citeauthoryear{Otterbacher, Checco, Demartini, and
  Clough}{Otterbacher et~al\mbox{.}}{2018}]%
        {otterbacher2018investigating}
\bibfield{author}{\bibinfo{person}{Jahna Otterbacher},
  \bibinfo{person}{Alessandro Checco}, \bibinfo{person}{Gianluca Demartini},
  {and} \bibinfo{person}{Paul Clough}.} \bibinfo{year}{2018}\natexlab{}.
\newblock \showarticletitle{Investigating user perception of gender bias in
  image search: the role of sexism}. In \bibinfo{booktitle}{\emph{The 41st
  International ACM SIGIR Conference on Research \& Development in Information
  Retrieval}}. \bibinfo{pages}{933--936}.
\newblock


\bibitem[\protect\citeauthoryear{Pan, Hembrooke, Joachims, Lorigo, Gay, and
  Granka}{Pan et~al\mbox{.}}{2007}]%
        {pan2007google}
\bibfield{author}{\bibinfo{person}{Bing Pan}, \bibinfo{person}{Helene
  Hembrooke}, \bibinfo{person}{Thorsten Joachims}, \bibinfo{person}{Lori
  Lorigo}, \bibinfo{person}{Geri Gay}, {and} \bibinfo{person}{Laura Granka}.}
  \bibinfo{year}{2007}\natexlab{}.
\newblock \showarticletitle{In Google we trust: Users’ decisions on rank,
  position, and relevance}.
\newblock \bibinfo{journal}{\emph{Journal of Computer-Mediated Communication}}
  (\bibinfo{year}{2007}), \bibinfo{pages}{801--823}.
\newblock


\bibitem[\protect\citeauthoryear{Pang, Lan, Guo, Xu, Wan, and Cheng}{Pang
  et~al\mbox{.}}{2016}]%
        {pang2016text}
\bibfield{author}{\bibinfo{person}{Liang Pang}, \bibinfo{person}{Yanyan Lan},
  \bibinfo{person}{Jiafeng Guo}, \bibinfo{person}{Jun Xu},
  \bibinfo{person}{Shengxian Wan}, {and} \bibinfo{person}{Xueqi Cheng}.}
  \bibinfo{year}{2016}\natexlab{}.
\newblock \showarticletitle{Text matching as image recognition}. In
  \bibinfo{booktitle}{\emph{Proceedings of the AAAI Conference on Artificial
  Intelligence}}.
\newblock


\bibitem[\protect\citeauthoryear{Rekabsaz and Schedl}{Rekabsaz and
  Schedl}{2020}]%
        {rekabsaz2020neural}
\bibfield{author}{\bibinfo{person}{Navid Rekabsaz} {and}
  \bibinfo{person}{Markus Schedl}.} \bibinfo{year}{2020}\natexlab{}.
\newblock \showarticletitle{Do Neural Ranking Models Intensify Gender Bias?}.
  In \bibinfo{booktitle}{\emph{Proceedings of the 43rd International ACM SIGIR
  Conference on Research and Development in Information Retrieval}}.
  \bibinfo{pages}{2065--2068}.
\newblock


\bibitem[\protect\citeauthoryear{Rekabsaz, West, Henderson, and
  Hanbury}{Rekabsaz et~al\mbox{.}}{2021}]%
        {rekabsaz2018measuring}
\bibfield{author}{\bibinfo{person}{Navid Rekabsaz}, \bibinfo{person}{Robert
  West}, \bibinfo{person}{James Henderson}, {and} \bibinfo{person}{Allan
  Hanbury}.} \bibinfo{year}{2021}\natexlab{}.
\newblock \showarticletitle{Measuring Societal Biases in Text Corpora via
  First-Order Co-occurrence}.
\newblock \bibinfo{journal}{\emph{Proceedings of the International AAAI
  Conference on Web and Social Media (ICWSM)}} (\bibinfo{year}{2021}).
\newblock


\bibitem[\protect\citeauthoryear{Robertson and Zaragoza}{Robertson and
  Zaragoza}{2009}]%
        {robertson2009probabilistic}
\bibfield{author}{\bibinfo{person}{Stephen Robertson} {and}
  \bibinfo{person}{Hugo Zaragoza}.} \bibinfo{year}{2009}\natexlab{}.
\newblock \showarticletitle{The probabilistic relevance framework: BM25 and
  beyond}.
\newblock \bibinfo{journal}{\emph{Foundations and Trends{\textregistered} in
  IR}} (\bibinfo{year}{2009}).
\newblock


\bibitem[\protect\citeauthoryear{Romanov, De-Arteaga, Wallach, Chayes, Borgs,
  Chouldechova, Geyik, Kenthapadi, Rumshisky, and Kalai}{Romanov
  et~al\mbox{.}}{2019}]%
        {romanov2019s}
\bibfield{author}{\bibinfo{person}{Alexey Romanov}, \bibinfo{person}{Maria
  De-Arteaga}, \bibinfo{person}{Hanna Wallach}, \bibinfo{person}{Jennifer
  Chayes}, \bibinfo{person}{Christian Borgs}, \bibinfo{person}{Alexandra
  Chouldechova}, \bibinfo{person}{Sahin Geyik}, \bibinfo{person}{Krishnaram
  Kenthapadi}, \bibinfo{person}{Anna Rumshisky}, {and} \bibinfo{person}{Adam
  Kalai}.} \bibinfo{year}{2019}\natexlab{}.
\newblock \showarticletitle{What’s in a Name? Reducing Bias in Bios without
  Access to Protected Attributes}. In \bibinfo{booktitle}{\emph{Proceedings of
  the Conference of the North American Chapter of the Association for
  Computational Linguistics: Human Language Technologies}}.
  \bibinfo{pages}{4187--4195}.
\newblock


\bibitem[\protect\citeauthoryear{Singh and Joachims}{Singh and
  Joachims}{2018}]%
        {singh2018fairness}
\bibfield{author}{\bibinfo{person}{Ashudeep Singh} {and}
  \bibinfo{person}{Thorsten Joachims}.} \bibinfo{year}{2018}\natexlab{}.
\newblock \showarticletitle{Fairness of exposure in rankings}. In
  \bibinfo{booktitle}{\emph{Proceedings of SIGKDD}}.
\newblock


\bibitem[\protect\citeauthoryear{Singh and Joachims}{Singh and
  Joachims}{2019}]%
        {singh2019policy}
\bibfield{author}{\bibinfo{person}{Ashudeep Singh} {and}
  \bibinfo{person}{Thorsten Joachims}.} \bibinfo{year}{2019}\natexlab{}.
\newblock \showarticletitle{Policy Learning for Fairness in Ranking}. In
  \bibinfo{booktitle}{\emph{Conference on Neural Information Processing Systems
  (NeurIPS)}}.
\newblock


\bibitem[\protect\citeauthoryear{Sparck~Jones and Van~Rijsbergen}{Sparck~Jones
  and Van~Rijsbergen}{1975}]%
        {sparck1975report}
\bibfield{author}{\bibinfo{person}{K Sparck~Jones} {and} \bibinfo{person}{C
  Van~Rijsbergen}.} \bibinfo{year}{1975}\natexlab{}.
\newblock \showarticletitle{Report on the Need for and Provision of an
  'ideal’ information retrieval test collection}.
\newblock \bibinfo{journal}{\emph{British Library Research and Development
  Report 5266}} (\bibinfo{year}{1975}).
\newblock


\bibitem[\protect\citeauthoryear{S{\"u}hr, Biega, Zehlike, Gummadi, and
  Chakraborty}{S{\"u}hr et~al\mbox{.}}{2019}]%
        {suhr2019two}
\bibfield{author}{\bibinfo{person}{Tom S{\"u}hr}, \bibinfo{person}{Asia~J
  Biega}, \bibinfo{person}{Meike Zehlike}, \bibinfo{person}{Krishna~P Gummadi},
  {and} \bibinfo{person}{Abhijnan Chakraborty}.}
  \bibinfo{year}{2019}\natexlab{}.
\newblock \showarticletitle{Two-sided fairness for repeated matchings in
  two-sided markets: A case study of a ride-hailing platform}. In
  \bibinfo{booktitle}{\emph{Proceedings of the 25th ACM SIGKDD International
  Conference on Knowledge Discovery \& Data Mining}}.
  \bibinfo{pages}{3082--3092}.
\newblock


\bibitem[\protect\citeauthoryear{Turc, Chang, Lee, and Toutanova}{Turc
  et~al\mbox{.}}{2019}]%
        {turc2019wellread}
\bibfield{author}{\bibinfo{person}{Iulia Turc}, \bibinfo{person}{Ming-Wei
  Chang}, \bibinfo{person}{Kenton Lee}, {and} \bibinfo{person}{Kristina
  Toutanova}.} \bibinfo{year}{2019}\natexlab{}.
\newblock \bibinfo{title}{Well-Read Students Learn Better: On the Importance of
  Pre-training Compact Models}.
\newblock
\newblock
\showeprint[arxiv]{1908.08962}~[cs.CL]


\bibitem[\protect\citeauthoryear{Wienclaw}{Wienclaw}{2011}]%
        {wienclaw2011gender}
\bibfield{author}{\bibinfo{person}{Ruth~A Wienclaw}.}
  \bibinfo{year}{2011}\natexlab{}.
\newblock \showarticletitle{Gender roles}.
\newblock \bibinfo{journal}{\emph{Sociology Reference Guide: Gender Roles and
  Equality}} (\bibinfo{year}{2011}), \bibinfo{pages}{33--40}.
\newblock


\bibitem[\protect\citeauthoryear{Xie, Dai, Du, Hovy, and Neubig}{Xie
  et~al\mbox{.}}{2017}]%
        {xie2017controllable}
\bibfield{author}{\bibinfo{person}{Qizhe Xie}, \bibinfo{person}{Zihang Dai},
  \bibinfo{person}{Yulun Du}, \bibinfo{person}{Eduard Hovy}, {and}
  \bibinfo{person}{Graham Neubig}.} \bibinfo{year}{2017}\natexlab{}.
\newblock \showarticletitle{Controllable invariance through adversarial feature
  learning}. In \bibinfo{booktitle}{\emph{Proceedings of the 31st International
  Conference on Neural Information Processing Systems}}.
  \bibinfo{pages}{585--596}.
\newblock


\bibitem[\protect\citeauthoryear{Xiong, Dai, Callan, Liu, and Power}{Xiong
  et~al\mbox{.}}{2017}]%
        {xiong2017end}
\bibfield{author}{\bibinfo{person}{Chenyan Xiong}, \bibinfo{person}{Zhuyun
  Dai}, \bibinfo{person}{Jamie Callan}, \bibinfo{person}{Zhiyuan Liu}, {and}
  \bibinfo{person}{Russell Power}.} \bibinfo{year}{2017}\natexlab{}.
\newblock \showarticletitle{End-to-end neural ad-hoc ranking with kernel
  pooling}. In \bibinfo{booktitle}{\emph{Proceedings of the 40th International
  ACM SIGIR Conference on Research and Development in Information Retrieval}}.
  \bibinfo{pages}{55--64}.
\newblock


\bibitem[\protect\citeauthoryear{Xiong, Xiong, Li, Tang, Liu, Bennett, Ahmed,
  and Overwikj}{Xiong et~al\mbox{.}}{2021}]%
        {xiong2021approximate}
\bibfield{author}{\bibinfo{person}{Lee Xiong}, \bibinfo{person}{Chenyan Xiong},
  \bibinfo{person}{Ye Li}, \bibinfo{person}{Kwok-Fung Tang},
  \bibinfo{person}{Jialin Liu}, \bibinfo{person}{Paul~N. Bennett},
  \bibinfo{person}{Junaid Ahmed}, {and} \bibinfo{person}{Arnold Overwikj}.}
  \bibinfo{year}{2021}\natexlab{}.
\newblock \showarticletitle{Approximate Nearest Neighbor Negative Contrastive
  Learning for Dense Text Retrieval}. In \bibinfo{booktitle}{\emph{Proceedings
  of the International Conference on Learning Representations}}.
\newblock


\bibitem[\protect\citeauthoryear{Yang and Stoyanovich}{Yang and
  Stoyanovich}{2017}]%
        {yang2017measuring}
\bibfield{author}{\bibinfo{person}{Ke Yang} {and} \bibinfo{person}{Julia
  Stoyanovich}.} \bibinfo{year}{2017}\natexlab{}.
\newblock \showarticletitle{Measuring fairness in ranked outputs}. In
  \bibinfo{booktitle}{\emph{Proceedings of Conference on Scientific and
  Statistical Database Management}}.
\newblock


\bibitem[\protect\citeauthoryear{Yang, Fang, and Lin}{Yang
  et~al\mbox{.}}{2017}]%
        {yang2017anserini}
\bibfield{author}{\bibinfo{person}{Peilin Yang}, \bibinfo{person}{Hui Fang},
  {and} \bibinfo{person}{Jimmy Lin}.} \bibinfo{year}{2017}\natexlab{}.
\newblock \showarticletitle{Anserini: Enabling the use of Lucene for
  information retrieval research}. In \bibinfo{booktitle}{\emph{Proceedings of
  the 40th International ACM SIGIR Conference on Research and Development in
  Information Retrieval}}. \bibinfo{pages}{1253--1256}.
\newblock


\bibitem[\protect\citeauthoryear{Zehlike, Bonchi, Castillo, Hajian, Megahed,
  and Baeza-Yates}{Zehlike et~al\mbox{.}}{2017}]%
        {zehlike2017fa}
\bibfield{author}{\bibinfo{person}{Meike Zehlike}, \bibinfo{person}{Francesco
  Bonchi}, \bibinfo{person}{Carlos Castillo}, \bibinfo{person}{Sara Hajian},
  \bibinfo{person}{Mohamed Megahed}, {and} \bibinfo{person}{Ricardo
  Baeza-Yates}.} \bibinfo{year}{2017}\natexlab{}.
\newblock \showarticletitle{Fa* ir: A fair top-k ranking algorithm}. In
  \bibinfo{booktitle}{\emph{Proceedings of the 2017 ACM on Conference on
  Information and Knowledge Management}}. \bibinfo{pages}{1569--1578}.
\newblock


\bibitem[\protect\citeauthoryear{Zehlike and Castillo}{Zehlike and
  Castillo}{2020}]%
        {zehlike2020reducing}
\bibfield{author}{\bibinfo{person}{Meike Zehlike} {and} \bibinfo{person}{Carlos
  Castillo}.} \bibinfo{year}{2020}\natexlab{}.
\newblock \showarticletitle{Reducing disparate exposure in ranking: A learning
  to rank approach}. In \bibinfo{booktitle}{\emph{Proceedings of The Web
  Conference}}. \bibinfo{pages}{2849--2855}.
\newblock


\bibitem[\protect\citeauthoryear{Zemel, Wu, Swersky, Pitassi, and Dwork}{Zemel
  et~al\mbox{.}}{2013}]%
        {zemel2013learning}
\bibfield{author}{\bibinfo{person}{Rich Zemel}, \bibinfo{person}{Yu Wu},
  \bibinfo{person}{Kevin Swersky}, \bibinfo{person}{Toni Pitassi}, {and}
  \bibinfo{person}{Cynthia Dwork}.} \bibinfo{year}{2013}\natexlab{}.
\newblock \showarticletitle{Learning fair representations}. In
  \bibinfo{booktitle}{\emph{Proceedings of the International conference on
  Machine Learning}}. PMLR, \bibinfo{pages}{325--333}.
\newblock


\bibitem[\protect\citeauthoryear{Zhang, Lemoine, and Mitchell}{Zhang
  et~al\mbox{.}}{2018}]%
        {zhang2018mitigating}
\bibfield{author}{\bibinfo{person}{Brian~Hu Zhang}, \bibinfo{person}{Blake
  Lemoine}, {and} \bibinfo{person}{Margaret Mitchell}.}
  \bibinfo{year}{2018}\natexlab{}.
\newblock \showarticletitle{Mitigating unwanted biases with adversarial
  learning}. In \bibinfo{booktitle}{\emph{Proceedings of the 2018 AAAI/ACM
  Conference on AI, Ethics, and Society}}. \bibinfo{pages}{335--340}.
\newblock


\bibitem[\protect\citeauthoryear{Zobel}{Zobel}{1998}]%
        {zobel1998reliable}
\bibfield{author}{\bibinfo{person}{Justin Zobel}.}
  \bibinfo{year}{1998}\natexlab{}.
\newblock \showarticletitle{How reliable are the results of large-scale
  information retrieval experiments?}. In \bibinfo{booktitle}{\emph{Proceedings
  of the 21st International ACM SIGIR Conference on Research and Development in
  Information Retrieval}}. \bibinfo{pages}{307--314}.
\newblock


\end{thebibliography}

\end{document}